\documentstyle[aps,epsf,preprint]{revtex}
\begin{document}
\draft 
\preprint{Submitted to Physical Review B}
\title{Phase Transitions in the Two-Dimensional {\it XY} Model\\ with 
Random Phases : a Monte Carlo Study.}

\author{J. Maucourt, and D.R. Grempel}
\address{CEA/D\'{e}partement de Recherche Fondamentale sur la 
Mati\`{e}re 
Condens\'{e}e.\\
SPSMS, CENG, 17, rue des Martyrs, 38064, Grenoble Cedex 9, 
France.}

\date{\today}

\maketitle

\begin{abstract}
We study the two-dimensional {\it XY} model with quenched random 
phases by Monte Carlo simulation and finite-size scaling analysis. We determine the phase diagram of the model and study its critical behavior as a function of disorder and  temperature.  If the strength of the randomness is less than a critical value, $\sigma_{c}$, the system has a Kosterlitz-Thouless (KT) phase transition from the paramagnetic phase to a state with quasi-long-range order. Our data suggest that the latter  exists  down to $T=0$ in contradiction with theories that predict the appearance of 
a low-temperature reentrant phase.  At the critical disorder $T_{KT}\rightarrow 0$ and for $\sigma > \sigma_{c}$ there is no quasi-ordered phase. At zero temperature there is a phase transition between two different glassy states at $\sigma_{c}$. The functional dependence of the correlation length  on $\sigma$ suggests that this transition corresponds to the disorder-driven unbinding of vortex pairs.
\end{abstract}

\pacs{PACS numbers: 75.10.Nr, 75.40.Mg, 74.40}

\narrowtext

\section{introduction}
\label{sec:introduction}

The two-dimensional {\it XY} model with quenched phase disorder 
describes
 the thermodynamics of a variety of systems of experimental 
interest.  
Examples of these include magnetic systems with 
random Dzyaloshinkii-Moriya interactions \cite{Rubinstein}, crystal 
systems on 
disordered substrates \cite{ChaI},  arrays of Josephson junctions 
with 
positional disorder \cite{GranatoI,GranatoII} and vortex 
glasses \cite{Fisher}. The critical properties of these systems are 
described by the  model Hamiltonian 

\begin{eqnarray}
	H & = & -J\sum_{<i,j>}\cos 
	\left(\theta_{i}-\theta_{j}-A_{ij}\right),
	\label{hamiltonian}
\end{eqnarray}

\noindent  where the sum runs over the $N_{b}$ bonds of a $2-d$ 
square 
lattice, $\theta_{i}$ is a phase variable attached to site $i$  
and $A_{ij}=-A_{ji}$ is a quenched random bond-variable whose precise 
physical meaning depends on the nature of the system 
described by the model. We assume for simplicity that the phase 
shifts on different bonds 
are uncorrelated and that each configuration $\{A_{ij}\}$  
occurs with  probability 

\begin{eqnarray}
	P\left[A\right]=\frac{1}{\left(2\pi\sigma\right)^{N_{b}/2}}\exp\left[-\frac{1}{2\sigma}
	\sum_{<ij>} A_{ij}^{2}\right].
	\label{probability}
\end{eqnarray}

The nature of the phase diagram of this system in the $\sigma-T$ 
plane 
has been controversial. By mapping model (\ref{hamiltonian}) into a 
two-dimensional Coulomb
 gas in a random background of frozen dipoles,  Rubinstein 
{\it et al.} \cite{Rubinstein} generalized 
the original treatment \cite{Kosterlitz} of the Kosterlitz-Thouless 
(KT) transition to the  random case. Their analysis leads to the 
phase diagram shown in 
Fig.\ \ref{fig-phasediagram}. 
For $\sigma>\sigma_{c}=\pi/8$ the system is 
paramagnetic at all temperatures. For  $\sigma<\sigma_{c}$, in the 
limit 
of infinite vortex core energy ,  two lines,  
 $T_{\pm}(\sigma)=\frac{\pi J}{4}\left[1\pm\sqrt{1-8\sigma/\pi}\right]$,
define
the boundaries of a quasi-ordered KT phase. The transition along the 
upper curve is similar to the KT transition of the pure system 
except that the transition temperature and the value of the jump of 
the exponent 
$\eta$ at the transition
are reduced by the randomness. The second transition line has no 
counterpart in the pure 
case and it signals the reentrance of the disordered phase at low 
temperature. It occurs because of the presence of a term 
in the renormalization-group flow 
equations\cite{Rubinstein} that makes the 
 vortex fugacity  grow  continuously with increasing length scale at 
low temperature. 
The effect of a finite vortex-core energy is to push the two 
transitions down to lower 
temperatures, 
 $T_{KT}(\sigma)<T_{+}(\sigma)$,  $T_{r}(\sigma)<T_{-}(\sigma)$.

So far, the reentrant phase has not been seen either in numerical
 simulations\cite{ForresterI,Chakrabarti} or  experiments 
\cite{ForresterII} 
on disordered arrays of Josephson junctions. Although finite-size 
\cite{ForresterI} 
or pinning \cite{ForresterII} effects can account for the  failure 
to observe the reentrance transition, it has  been recently
 suggested \cite{ChaII,Natterman} that, in fact, the latter does not 
exist. The physical argument supporting this claim is that 
topological defects can appear 
 at $T=0$ only  if the energy cost to 
create them is balanced by an energy gain coming from the random 
field. 
However, a simple probabilistic estimate  \cite{ChaII} shows that, 
in an infinite system, the probability of finding sites  
on which it 
pays to create an isolated vortex vanishes below $\sigma_{c}$. On the 
basis of this   
reasoning one expects the quasi-ordered phase to be stable below 
the critical disorder at $T=0$. 
Refining this heuristic considerations by taking into account the 
interactions 
 between defects, Natterman {\it et al.} \cite{Natterman} have 
derived 
 a set of renormalization group 
 equations that reduce to those 
derived earlier\cite{Rubinstein} above a crossover temperature
 $T^{\ast}\approx 2J\sigma$, but are different from them below it. 
 In the modified theory, vortices are irrelevant  
 in the whole region $\sigma< \sigma_{c}$, $T<T_{KT}(\sigma)$,  
 where the correlation functions exhibit power-law decay. At  $T=0$ 
 they find a disordered-driven transition at 
 $\sigma=\sigma_{c}$ at which there is a jump of  the correlation
  function exponent from a finite value to zero with $\Delta 
\eta_{c}=1/16$.
   Above $\sigma_{c}$ the correlation length is finite.

 In this paper we present the results 
 of extensive Monte Carlo  simulations of 
 the {\it XY} model with random phases. The main difference between 
 our simulations and  those performed previously
 \cite{ForresterI,Chakrabarti} resides in the use of finite-size 
scaling in the  
 analysis of the results. Considerations of scale invariance and 
universality 
 provide a sensitive tool to distinguish and characterize  the 
 different possible phases by imposing stringent conditions on the 
form of the correlation functions as functions temperature, 
randomness, and system-size.

Scaling behaviour in the paramagnetic phase, near the KT transition, 
is most conveniently discussed in terms of the 
Binder function\cite{Binder}, essentially a ratio of moments of the 
order parameter. Parametrizing the correlation length in the KT form, 
$\xi(T)\sim \exp A/\sqrt{T/T_{KT}-1}$, and 
adjusting $A$ and $T_{KT}$ so  
that scaling holds, we determined the $\sigma$-dependence 
of the  KT transition temperature and   
of the 
expected \cite{Rubinstein} reentrant transition temperature.  
We found that $T_{KT}(\sigma)$ vanishes sharply at 
$\sigma_{c}=\pi/8$, 
in agreement with a prediction of Ozeki and Nishimori \cite{Ozeki}.  
 The measured 
disorder-dependence of $T_{KT}$ near the critical disorder is 
consistent with the expression that can be 
derived from the modified renormalization group 
equations\cite{Natterman}. We found no 
trace of the expected signature (cf. Section \ref{sec:method:scaling})
of a reentrant transition in the size dependence of the Binder 
function at low temperatures. 

For $T\le T_{KT}$, analysis of the size-dependence 
of the moments of the order parameter gives access to the 
temperature dependence of the correlation function exponent, 
$\eta(T,\sigma)$. It can be shown that, in the case of a reentrant transition, $d\eta/dT$ must change sign going from positive near 
$T_{KT}$ to negative near $T_{r}$. 
 We found instead that $\eta(T)$
  is  monotonic and assymptotically approaches 
 the spinwave result \cite{Natterman} $\eta(T)\sim 
\frac{1}{2\pi}(T/J+\sigma)$
  at low temperature. This suggests that the quasi-ordered phase is 
stable down to $T=0$ for 
sufficiently weak disorder. 

A scaling analysis of the type described above 
 is not possible above $\sigma_{c}$ because in this region the 
disorder and temperature 
dependence of the correlation length is 
not known {\it a priori}. However, from the analysis of the data in 
the weak-disorder region one can determine not only the critical 
temperature and exponents but also
the full shape of the scaling function. Since by universality the 
latter is the same 
throughout the paramagnetic region  , 
we can use it to analize the resuls for $\sigma\ge\sigma_{c}$ as well. 
We found that in the strong disorder region  the spin-spin 
correlation function decays exponentially at all 
temperatures. The correlation length decreases with increasing $T$ 
and stays finite 
as $T\rightarrow 0$.  The extrapolation of our results to $T=0$ is 
consistent 
with the form $\log \xi(T=0,\sigma) \sim (\sigma-\sigma_{c})^{-1}$ 
expected near a disorder-driven vortex-unbinding 
transition\cite{Natterman}.

\section{METHOD}
\label{sec:method}

In this Section we describe the technique and the 
method  
of data analysis that we have 
used in our simulations. 

\subsection{Scaling Analysis.}  
\label {sec:method:scaling}

The analysis is based on the finite-size scaling  properties of the 
moments of the order 
parameter, 

\begin{eqnarray}
q^{(n)}=\left[\left<\left(\frac{1}{N}\sum_{i=1}^{N}\cos\theta_{i}\right)^{n}
\right>\right]_{d}, 
\label{moments}
\end{eqnarray}

\noindent where $N=L^{2}$ is the total number of spins in the system 
and $L$ is its linear dimension. Here 
the symbol $\left<\ldots\right>$ indicates  
the usual thermodynamic average with the Gibbs measure,  
$P\sim \exp(-\beta H)$, and 
$\left[\ldots\right]_{d}$ is the average over the phase shift 
distribution
 given in Eq.\ \ref{probability}.

In a finite system all configurations related 
by a global spin rotation enter in the thermodynamic average 
with equal weight. Since the  
average over the global rotation angle of all odd moments vanishes,  
the  first non-trivial moments 
are  $q^{(2)}$ and $q^{(4)}$.  If $L\gg a$ ($a$ is the 
lattice spacing), we may write

\begin{eqnarray}
	\begin{array}{l}
	 q^{(2)}(L,T,\sigma)=L^{-2}\int_{{\cal S}}d^{2}r\ 
	 C_{2}(r), \\
	q^{(4)}(L,T,\sigma)=L^{-6}\int_{{\cal S}}
	 d^{2}r_{1}d^{2}r_{2}d^{2}r_{3}\ C_{4}(r_{1},r_{2},r_{3}), 
	\end{array}
\label{2nd4thmoments}
\end{eqnarray}

\noindent where the integrals are over the surface of the system and 
$C_{n}$, $n=2,4$,  
are the two and four-point 
correlation functions respectively. These are given by the 
rotationally invariant expressions 

\begin{eqnarray}
	\begin{array}{c}
	C_{2}(r)=\frac{1}{2}\ \left[\left<\cos 
	[\theta(\vec{r})-\theta(0)]\right>\right]_{d}, \\  
	C_{4}(\vec{r}_{1},\vec{r}_{2},\vec{r}_{3})=\frac{3}{8}\  
	\left[\left<\cos 
	[\theta(\vec{r}_{1})-\theta(\vec{r}_{2})+\theta(\vec{r}_{3})-
	\theta(0)]\right>\right]_{d}. 
	\end{array}
\label{2nd4thcorrelations}
\end{eqnarray}

The scaling theory of phase transitions may be used to determine the 
general form  of the moments as functions 
of temperature, randomness and system 
size.  In the
paramagnetic phase, sufficiently close to $T_{KT}$ for the 
system to be in the scaling regime, the
 two-point function scales as \cite{Binder}

\begin{eqnarray}
	C_{2}(\vec{r})\sim r^{-\eta}\tilde{C}_{2}(r/\xi), 
	\label{scaling2ndcorrelation}
\end{eqnarray}

\noindent where $\eta$ is the correlation function exponent, 
$\xi\equiv\xi(\sigma,T)$ 
is the correlation length,  and the function $\tilde{C}_{2}(x)$ is 
universal. 
Using Eq.\ \ref{scaling2ndcorrelation} and the corresponding 
expression for the 
four-point function, $q^{(2)}$ 
 and $q^{(4)}$ may be written in the form :

\begin{eqnarray}
\begin{array}{l}
	q^{(2)}(L,T,\sigma)=L^{-\eta}Q_{+}^{(2)}(L/\xi), \\
	q^{(4)}(L,T,\sigma)=L^{-2\eta}Q_{+}^{(4)}(L/\xi),
\end{array}
\label{scalingof moments}	
\end{eqnarray}

\noindent valid in the disordered phase. The functions $Q_{+}^{(2)}$ 
and 
$Q_{+}^{(4)}$ in the expressions above are 
 universal functions of their argument. 
 
The correlation length diverges at $T_{KT}$ and stays infinite 
throughout the KT phase. 
In this phase the long-wavelength behavior of the system is 
described by renormalized spinwave theory. The two and four-point 
functions may be easily calculated within this theory resulting in 
the expressions : 

\begin{eqnarray}
	\begin{array}{l}
	 C_{2}(r)\sim r^{-\eta}, \\
	 C_{4}(\vec{r}_{1},\vec{r}_{2},\vec{r}_{3})
	 \sim \left[\frac{r_{2}\ |\vec{r}_{1}-\vec{r}_{3}|}
	 {r_{1}\ r_{3}\ |\vec{r}_{1}-\vec{r}_{2}|\ 
	 |\vec{r}_{2}-\vec{r}_{3}|}\right]^{\eta}.
	\end{array}
\label{correlationsKT}
\end{eqnarray}

Substituting  Eq.\ \ref{correlationsKT} into Eq.\ \ref{2nd4thmoments} 
we obtain the scaling form of the moments in the quasi-ordered KT 
phase  :

\begin{eqnarray}
	\begin{array}{l}
	q^{(2)}(r,T,\sigma)\sim L^{-\eta}\ 
	Q_{-}^{(2)}(\eta), \\
	 q^{(4)}(r,T,\sigma)\sim L^{-2\eta}
	 \ Q_{-}^{(4)}(\eta),  
	\end{array} 
\label{momentsKT}
\end{eqnarray}

\noindent where the universal functions $Q_{-}^{(2)}$ and 
$Q_{-}^{(4)}$, given by
\begin{eqnarray}
	\begin{array}{c}
	Q_{-}^{(2)}(\eta)\sim \int_{} d^{2}x |x|^{-\eta},\\
	 Q_{-}^{(4)}(\eta)=\int_{}
	 d^{2}x_{1}d^{2}x_{2}d^{2}x_{3}
	 \left[\frac{x_{2}\ |\vec{x}_{1}-\vec{x}_{3}|}
	 {x_{1}\ x_{3}\ |\vec{x}_{1}-\vec{x}_{2}|\ 
	 |\vec{x}_{2}-\vec{x}_{3}|}\right]^{\eta}, 
	\end{array}
\label{univlowT}
\end{eqnarray}

\noindent depend on temperature and disorder only through the 
correlation function exponent $\eta$.

In the first step of the analysis the unknown 
factor  $L^{-\eta}$ can be eliminated from  Eqs.\ \ref{scalingof 
moments} and \ref{momentsKT} 
 by working with the ratio $u(L,T)=q^{(4)}/[q^{(2)}]^{2}$. It is customary to normalize this quantity  so  
that it vanishes at high 
temperature and it goes to unity for a homogeneous ferromagnet at $T=0$. 
The normalized quantity, known as the  Binder function\cite{Binder}, is defined by

\begin{eqnarray}
	g(L,T)=\frac{2}{3}(3-u(L,T)), 
	\label{binderfunction}
\end{eqnarray}

\noindent It follows from  Eqs.\ \ref{scalingof 
moments} and \ref{momentsKT} that the Binder 
function obeys the scaling law

\begin{eqnarray}
	g(L,T,\sigma)=\left\{
	\begin{array}{ll}
		G_{+}(L/\xi)
		&,T\ge T_{KT}, \\
		G_{-}(\eta)
		&,T\le T_{KT}, 
	\end{array}
	\right.
	\label{scalingbinder}
\end{eqnarray}

\noindent where $G_{\pm}$ are the universal scaling functions in the paramagnetic and quasi-ordered phases, respectively. Our discussion in the Section below is 
based on the following consequences of Eq.\ \ref{scalingbinder} :  
\begin{enumerate}
\item In the paramagnetic phase $g(L,T)$ depends on both temperature 
and system size. 
However, by choosing  $\xi(T,\sigma)$ appropriately one can make all  
the data  collapse on a  universal function, $G_{+}(L/\xi)$.
\item At the KT transition temperature and in the KT phase $\xi$ is infinite. 
Therefore the curves for different 
sizes must merge at $T_{KT}$. In the KT phase the Binder function 
depends on temperature and disorder through a universal function of the correlation function exponent, 
$G_{-}(\eta)$.
 \item A logarithmic plot 
of $q^{(2k)}(L,T)$ as a function of $L$ for different temperatures in 
the $KT$ phase must give a series of straight lines whose common slope is $-k\eta$. 
\item If a  reentrant transition occurs,  the curves $g(L,T)$ for the 
different sizes must split again at $T_r$ and remain distinct down to $T\rightarrow 
0$.
\end{enumerate}

In addition, one can show (see Section\ \ref{sec:results:weakdisorder:quasi-ordered phase}) that, if there is 
a reentrant transition, the slope $d\eta/dT$ must change sign at some temperature between $T_{KT}$ and $T_r$. This gives yet another signature of reentrance.

\subsection{Numerical method}
\label{sec:method:montecarlo}

We have determined by Monte Carlo simulation the $2-$nd and $4-$th 
moments of the order parameter 
for systems of planar spins 
on  an $L\times L$ square lattice with periodic boundary 
conditions. In studies of disordered systems it is particularly  
crucial to 
check that thermal equilibrium has been 
achieved before making the Monte Carlo measurements.  We have  done 
this using a procedure first introduced by  Bhatt et 
Young\cite{bhatt} 
for Ising glasses and generalized 
 by Ray and Moore\cite{ray} to the case of $XY$ spin glasses.  
 The method is based on the comparison between two quantities. One is the 
mean-square-averaged overlap (MSAO) of two time-delayed configurations of the same evolving system. The other one is the MSAO between the instantaneous configurations of two identical copies (or replicas) of the system that have evolved 
independently starting 
 from arbitrary 
initial conditions. The two overlaps \cite{ray} are defined by 

\begin{equation}
		{\cal O}(t+t_{w},t_{w}) = \left[\left<\left[\frac{1}{N}\sum_{i} 
	\cos\left[\theta_{i}(t+t_{w})-\theta_{i}(t_{w})\right]\right]^{2}\right>\right]_{d},  	
\end{equation}

\noindent and
\begin{equation}
		{\cal O}_{r}(t_{w}) =\left[\left<\left[\frac{1}{N}\sum_{i} 
		\cos\left[\theta^{(a)}_{i}(t_{w})-\theta^{(b)}_{i}(t_{w})\right]\right]^{2}\right>\right]_{d},   	
\end{equation}

\noindent respectively. Here,  $t_{w}$ is a waiting time during which 
the systems 
considered here have evolved from the initial conditions, and $t$ and  
$(a)$ and $(b)$ denote the time delay and the replicas referred to 
above. 
Let $t_{eq}$ and $t_{r}$ be the equilibration time and the 
equilibrium relaxation time, 
respectively. It follows from general considerations of ergodicity 
that ${\cal O}(t + 
t_{w},t_{w})\rightarrow{\cal O}_{r}(t_{w})$ when $t_{w}> t_{eq}$ and $t>t_{r}$.

In Monte Carlo simulations thermal averages are replaced by time 
averages. Thus, to compute the time-delayed overlap 
 the system is simulated for $t_{w}$ Monte Carlo time-steps per spin 
(MCS). The final configuration is stored, and the system is left to 
evolve further during  
 $t$  MCS. Measurements are taken at times 
 $t_{m}=t+t_{w}+m$, $m=0,\ldots,M$,  and the required overlap is 
 computed as the average of the overlaps  
 between the configurations at times $t_{m}$ and $t_{w}$. Since, in general, $t_{eq}\gg t_r$, we may take $t=t_{w}$ and 
write\cite{ray}  : 

\begin{eqnarray}
	{\cal O}(2 t_{w},t_{w}) \sim \left[\frac{1}{M} 
\sum_{m=1}^{M}\left[\frac{1}{N}\sum_{i} 
		\cos\left[\theta_{i}(2 
t_{w}+m)-\theta_{i}(t_{w})\right]\right]^{2}\right]_{d}.
	\label{overlapmc}
\end{eqnarray}

\noindent Similarly, in order to compute ${\cal O}_{r}(t_{w})$, we 
simulate in parallel two copies of 
the system for $t_{w}$ MCS steps and we take the average of the 
overlaps of the next  
$M$  configurations of the two replicas, 

\begin{eqnarray}
		{\cal 
O}_{r}(t_{w})\sim\left[\frac{1}{M}\sum_{m=1}^{M}\left[\frac{1}{N}\sum_{i} 
\cos\left[\theta^{(a)}_{i}(t_{w}+m)-\theta^{(b)}_{i}(t_{w}+m)\right]\right]^{2}\right]_{d}. 
\label{replicasmc}
\end{eqnarray}

In this work we have studied systems with $L=4,6,8,10,12$ and $16$  
for 
sixteen values of $\sigma$ in the range $0\le\sigma\le1$ and 
temperatures 
in the range $0.3\le T/J\le 
1.7$. For certain values of the disorder and for sizes $L\le10$ we 
pushed the simulations down to $T/J=0.1$. The thermalization times 
vary  
between $t_w=2\times 10^{3}$ for the smaller systems up to  $t_w=2\times 
10^{5}$ 
for the larger ones at low $T$. In all our simulations $M=t_w$. The number of realizations of 
the random bonds simulated  to perform the 
configuration average varies from $256$ for the $16\times 16$ systems 
up to 
$2048$ for the smallest ones. However, for values of $\sigma \sim 
\sigma_{c}$ we have averaged over 
four times us much bond configurations. Our simulations were 
performed on a $128-$processor CRAY T3D parallel computer.

\section{RESULTS}
\label{sec:results}

\subsection{Weak disorder}
\label{sec:results:weakdisorder}

We show in Figs.\ \ref{fig-1}a to \ref{fig-1}d the numerical values 
of the 
Binder function, Eq.\ \ref{binderfunction}, as a function of 
temperature  
and system-size for four 
values of $\sigma$,  representative of our results  below 
$\sigma_{c}\approx 0.393$. 
The four curves are  qualitatively
similar  except that the temperature scale shifts to the left with 
increasing disorder. 
For each size we can identify two regimes. At 
high temperatures $g(L,T)$ decreases with increasing temperature or 
system-size. 
There is an inflection point (not always seen in the 
temperature range shown in the figures) and, for sufficiently high
$T$,  $g(L,T)$  reaches 
a temperature-independent plateau 
whose hight depends on the system-size. 
At low temperatures $g(L,T)$ is a  convex function of $T$. Below a 
certain temperature the Binder function is $L$-independent within the 
statistical error. There is no further splitting 
of the curves at low $T$ in the temperature range covered by our simulations. 

These data follow closely the scenario  anticipated in the paragraph 
below Eq.\ \ref{scalingbinder} for a system that goes from 
a disordered to a quasi-ordered state through a Kosterlitz-Thouless 
transition 
at a temperature $T_{KT}$. The latter is identified as that at which the 
curves for different 
sizes merge. The observed absence of size-dependence of $g$ at low temperatures indicates
 that the domain of stability of the 
quasi-ordered phase 
extends, at least,  down to the lowest temperatures that we 
simulated.  It 
will be seen below that, except for the lowest concentrations, these 
are 
  below the expected\cite{Rubinstein} reentrance temperature. The 
  quantitative analysis of the data is as follows.  
 
\subsubsection{The paramagnetic region}
\label{sec:results:weakdisorder:paramagnetic region}

To discuss quantitatively the data in the high temperature region we 
assume
 that for each value of $\sigma$ the correlation length is of the KT 
form, 

\begin{equation}
\xi(T)\sim \exp A/\sqrt{T/T_{KT}-1},
\label{KTxi}
\end{equation}

\noindent where $A$ and $T/T_{KT}$ are disorder-dependent constants. 
These are determined such that for each $\sigma$ data for all 
temperatures 
and sizes collapse into a unique function of $L/\xi$ as required by 
Eq.\ \ref{scalingbinder}. 
Examples of the resulting scaling plots for $g(L,T)$ are shown in 
Fig.\ \ref{fig-2} for 
the same values of $\sigma$ as in Fig.\ \ref{fig-1}. It may be seen from the figure that the Binder function is a decreasing function of $L/\xi$. It decays exponentially for $L>>\xi$ and it varies lineraly for $L/\xi\sim 0$. At the transition temperature $g$ takes the universal value $g(L,T_{KT})=0.972\pm 0.001$. It will be seen below that the differences between the curves corresponding to the different values of $\sigma$ can be absorbed in a disorder-dependent amplitude in the correlation length, Eq.\ \ref{KTxi}.

The dependence of the critical 
temperature on randomness obtained from the scaling analysis is shown in Fig.\ 
\ref{fig-3}. For weak disorder $T_{KT}$ decreases slowly as a function of $\sigma$. When the 
transition temperature 
goes below $T^{\ast}$  its variation with $\sigma$
becomes steeper and it is seems to be very abrupt 
for $\sigma\approx\sigma_{c}$. In particular, whereas the system has a relatively high 
transition 
temperature for $\sigma=0.392$, the analysis of the shape of 
the Binder 
function (see below) shows that the $T=0$ correlation length is 
finite 
for $\sigma=0.393$. The observed sharpness of the transition line 
agrees with 
the prediction of Ozeki and 
Nishimori \cite{Ozeki} that, if a 
low temperature KT 
phase exists in this model,  the phase boundary must be 
parallel to the temperature axis as $T\rightarrow 0$. It may be shown\cite{Maucourt} that the modified RG equations of Natterman {\it et al.}\cite{Natterman} imply that, for 
$\sigma\approx \sigma_c$, 
$T_{KT}(\sigma)\approx 2 E_c/\ln\left[\pi^3/(\sigma_c-\sigma)\right]$ where $E_c$ is the vortex-core energy. We see from Fig.\ \ref{fig-3} that our results are consistent with this expression.
 
As a comparison of Figs.\ \ref{fig-3} and \ref{fig-phasediagram} shows,  
$T_{KT}$ is much reduced from its upper limit  $T_{+}$. This is an indication that the vortex-core energy is relatively small and may considerably renormalize the expected value of the reentrant transition temperature.
 The core energies can be computed from the measured $T_{KT}$ by  
 integrating the 
 renormalization group equations of Rubinstein {\it et al.}\cite{Rubinstein}  along the critical trajectory. The RG equations are :
 
 \begin{eqnarray}
 	\begin{array}{l}
 		\frac{dK}{dl}=-4\pi^{3} K^{2} y^{2},\\
 		\frac{dy}{dl}=(2-\pi K+\pi K^{2}\sigma)y,
 	\end{array}
 	\label{renormalizationgroup}
 \end{eqnarray}
 
 \noindent where $K=J/T$ is the running coupling constant and $y$ is 
the 
 vortex fugacity whose bare value is related to the temperature and 
 $E_{c}$ by $y_{0}(T)=\exp(-E_{c}/T)$.  The renormalization group 
trajectories, obtained by 
 integration of Eq.\ \ref{renormalizationgroup}, are given by 
 the family of curves
 
 \begin{eqnarray}
 	y^{2}=\frac{1}{2\pi^{3}} \left(2/K+\pi\ln K-\pi \sigma K\right)+C,
 	\label{trayectories}
 \end{eqnarray}
 
 \noindent where $C$ is an integration constant. The critical trajectory is 
determined by the condition that points 
 on it flow to the fixed point 
$y=0,K=J/T_{+}$ with $T_{+}=\frac{\pi J}{4}\left[1+\sqrt{1-8\sigma/\pi}\right]$.
The physical initial condition intersects the 
 critical trajectory at two temperatures, $T_{KT}$ and $T_{r}$, solutions of :
 
  \begin{equation}
 	\exp(-2E_c/T)=\frac{1}{2\pi^{3}} \left[2 (K^{-1}-K_{+}^{-1})+\pi\ln 
 	\frac{K}{K_{+}}-\pi \sigma (K-K_{+})\right].
 	\label{crit-tray}
 \end{equation}
 
\noindent The vortex-core energy $E_{c}$ may be determined by substituting the measured 
values of the KT
transition temperature in Eq.\ \ref{crit-tray}. We found that  
$E_{c}$ is a decreasing function of disorder that varies between 2.28$J$ for $\sigma=0$ and 1.9$J$ for $\sigma_{c}$. Once the vortex-core energy is known the
 reentrance temperature may easily be computed by searching for the second solution of Eq.\ \ref{crit-tray}. The result is shown by the open circles in 
Fig.\ \ref{fig-3}. It may be seen that, for $\sigma\ge0.25$, $T_r$ lies above the lowest measured 
temperatures. The fact that the Binder function shows no measurable size-dependence below $T_r$ for those  values of $\sigma$ for which the region $T<T_r$ is accessible to us suggests that for sufficiently  weak disorder the KT phase is stable 
at low temperature. The measurements of the disorder and temperature 
dependence of the exponent $\eta$ that we present below give further 
support to this conclusion.

\subsubsection{The quasi-ordered phase}
\label{sec:results:weakdisorder:quasi-ordered phase}

 The results in the KT phase below $T_{KT}$ are most conveniently 
analyzed in terms of the 
 scaling properties of the moments 
 of the order parameter, Eqs.\ \ref{momentsKT} and \ref{univlowT}. 
Although 
 all even moments contain the same information, we have worked with  
  $q^{(4)}$ because its temperature dependence is particularly simple. This is shown in  Fig.\ \ref{fig-5} where we plot raw data obtained 
for $\sigma=0.25$.  
 We see that $q^{(4)}$ is a decreasing function 
 of temperature and system size that can be quite accurately 
 fitted  by a straight line in the temperature range of the 
 simulations. According to Eq.\ \ref{momentsKT} we expect that all 
the 
 $L$-dependence 
 of $q^{(4)}$ be contained in the prefactor,  $L^{-\eta}$. The 
unknown $\eta$-dependence 
 in the function 
 $Q_{-}^{(4)}(\eta)$ may be eliminated from the problem by making 
 plots of  $\log q^{(4)}$ {\it vs. }$\log L$ for different values of the temperature and of the disorder. For a fixed value of $\sigma$ we obtain a series of straight lines, one for each temnperature, whose slopes are 
$-2 \eta$.  This  is illustrated in Fig.\ \ref{fig-7}, where we 
 represent the data for $\sigma=0.25$ as a function of $L$.
 
We determined by this procedure the  $T$-dependence 
 of $\eta$ for all values of $\sigma$. Results are plotted in Fig.\ 
\ref{fig-eta} for a few values of the disorder parameter.
 The correlation function exponent increases continuously with temperature. The extrapolation of our results down to $T=0$ is consistent with the spinwave 
 result, $\eta(T,\sigma)=(T/J+\sigma)/(2\pi)$. The correlation function exponent at the transition, $\eta(T_{KT})$, is disorder-dependent and smaller than $1/4$, the value for the pure system. The
 two sets of
 RG equations\cite{Rubinstein,Natterman} give different predictions 
for $\eta_c$. However, up to $\sigma=0.39$ the difference between the two theories is within our error bars and we can not decide in favour of one or the other.

The curves shown in Fig.\ \ref{fig-eta} are inconsistent 
 with the existence of the reentrant phase. Since all points on the critical trajectory flow to the same fixed point, the value of the correlation function exponent at criticality must be the same at $T_{r}$ and at 
 $T_{KT}$.  This implies that,  if there is a reentrant transition, $d\eta/dT$ must change sign at some temperature intermediate between $T_{KT}$ and $T_r$. Our data in the region $\sigma\ge 0.25$ where temperatures below $T_r$ are  accessible do not show this behaviour.

\subsection{Strong disorder.}
\label{sec:results:strongdisorder}

 The Binder function for four values of $\sigma >\sigma_{c}$ is 
shown in Figs.\ \ref{fig-8}a to \ref{fig-8}d. The behavior of $g(L,T)$ in these  cases is very 
different from that shown in Fig.\ \ref{fig-1} 
for weak 
disorder. The Binder functions for   
different sizes do not collapse 
 but they stay distinct down to the lowest
temperatures considered. Although it is obviously impossible to 
guarantee that they do not join at lower temperatures, this seems  
unlikely because, in order to merge, the curves would have to turn upwards at low 
temperatures. However, in all 
the cases in which we did observe a  KT transition, $g(L,T)$ was found 
to be a 
convex function of $T$ at low temperature.  We interpret this 
behavior as evidence that for $\sigma\ge\sigma_{c}$ the 
correlation length stays finite at 
all temperatures. 

For a fixed size,  
 $g(L,T)$ saturates to a constant value with decreasing $T$ indicating that, at low 
temperature, the 
correlation length only depends on the strength of the disorder. This 
dependence is very strong as shown in Fig.\ \ref{fig-9} where we 
represent 
  the temperature and $\sigma$  
dependence of $g(L,T)$ for $L=16$, our largest size, and $T/J\ge 0.3$. There is a five-fold decrease in the assymptotic value $g(L,T)$ when $\sigma$ varies between 0.4 and 1. 

In order to analyze our results quantitatively it is convenient to fit 
them to a simple analytic expression. We have chosen a four-parameter 
function,  

\begin{eqnarray}
	g(L,T)=m_{1}+m_{2}\left[1-\tanh[m_{3}(T-T_{0})]\right],
	\label{fitg}
\end{eqnarray}

\noindent that, despite its simplicity, accurately describes all the  
 data throughout  the paramagnetic phase as the solid lines in Figs.\ \ref{fig-8} and \ref{fig-9} show. Thanks to this 
parameterizatio our results can easily be extrapolated down to $T=0$. 
  
  The scaling analysis of the data in Figs.\ \ref{fig-8}a to 
  \ref{fig-8}d can not be performed in the same way as for 
  $\sigma\le\sigma_{c}$ because we do not have here an 
  explicit  expression giving the temperature dependence of the 
  correlation length. 
   However, this is not actually necessary.  By  
   universality, the scaling function $G_{+}(x)$ of Eq.\ \ref{scalingbinder} is the same 
  throughout the disordered phase. If we determine it from the data in 
the weak-disorder region,  we can use 
  it in the region $\sigma\ge\sigma_{c}$ too. 

In order to do this we reconsider the scaling plots of Fig.\ \ref{fig-2}.  The reason why the curves for different values of $\sigma$ are all different is 
 that in the scaling analysis the correlation length has been  
 determined up to 
 an amplitude that depends on the strength of 
 the disorder. We can compute all but one of the amplitudes by rescaling 
horizontally the  
 curves of Fig.\ \ref{fig-2} such that they can all be superimposed 
to one of them. Choosing the curve for $\sigma=0$ as reference and its amplitude as the unit of lenght, this procedure leads to the universal scaling function shown in Fig.\ \ref{fig-universal}. This plot where there more than six 
 hundred points corresponding to different system sizes, temperatures 
 and values of $\sigma$ lie on the same master curve is a beautiful example 
of universality. 

Having thus determined the scaling function $G_{+}(x)$y,  the correlation length in the strong-disorder regime can be computed  by reading from Fig.\ \ref{fig-universal} the values of $L/\xi$ corresponding to the measured values of $g(L,T)$. This yields the temperature and 
  disorder-dependent correlation length except for an undetermined 
  scale factor. The results of the analysis are presented in Fig.\ 
  \ref{fig-10}. At high temperature the correlation length is small and varies slowly with disorder. At low temperatures $\xi$ is $T$-independent and reaches a saturation value that increases very fast as we approach the critical disorder from the right. As seen in the figure, the error bars become very large when $\sigma\rightarrow \sigma_{c}$, the uncertainties in $G_{+}$ (cf. Fig.\ \ref{fig-universal}) and in the value of the Binder function leading to large
errors in $\xi$ for $L/\xi<<1$. The results can be reliably extrapolated to $T=0$ only for $\sigma\ge 0.49$. The disorder dependence of the ground-state correlation length in this region of values of $\sigma$ is represented in Fig.\ \ref{fig-11}. It can be seen that the  data are well described 
by the expression 
$\xi(T=0,\sigma)\sim\exp\left[b/(1-\sigma_{c}/\sigma)\right]$ that follows from the $T=0$ form of the modified RG equations of Natterman {\it et al.}\cite{Natterman}. This  
 result supports their conclusion that at zero-temperature there is a disorder-driven transition at $\sigma_{c}$ between two spin-glass states that differ by the properties of the spin-spin correlation function.   
  
In summary, we have studied the two-dimensional {\it XY} model 
with  random phases 
by Monte 
Carlo simulation. An essential element in our work is the  use 
of finite-size scaling in the analysis of the results. 
The scaling properties of the moments of the order parameter do not show the characteristic signatures that should be present if of a low-temperature reentrant transition occurred. Our results  suggest that renormalized spinwave theory is applicable as $T\rightarrow 0$ for sufficiently weak disorder. This is inconsistent with theories 
that predict the reentrance of the paramagnetic phase at low 
temperature\cite{Rubinstein} but agrees with more recent theories\cite{ChaII,Natterman}.

  \begin{figure}
           \caption{Schematic phase diagram of the {\it XY} model with random 
           phases in the $T-\sigma$ plane according to the theory of Rubinstein{\it et 
           al.}.  $T_{+}$ and 
           $T_{-}$ are, respectively,  
           the values of the KT transition temperature and of the 
           reentrance temperature for the case of an infinite vortex core energy.}
 	\label{fig-phasediagram}
 \end{figure}
 
 \begin{figure}
          \caption{The Binder function as a function of temperature 
          for different system sizes and several values of the 
          strength of the disorder in the weak disorder regime. (a)\  
$\sigma=0.09$, 
          (b)\ $\sigma=0.16$, (c)\ $\sigma=0.25$, and (d)\ $\sigma=0.36$.}
 	\label{fig-1}
 \end{figure}
 
 \begin{figure}
         \caption{A scaling plot of the data presented in the 
previous 
          figure. For each value of $\sigma$ the data for all 
          system-sizes and temperatures collapse into a single 
function of  
        $L/\xi(T,\sigma)$.}
         \label{fig-2}
 \end{figure}

 \begin{figure}
          \caption{The Kosterlitz-Thouless transition temperature as a
          function of disorder as  
          determined from the scaling plots. 
The dotted line is the
 prediction of the modified RG equations. 
The temperatures at which reentrance was expected to occur are indicated by the open circles.}
 \label{fig-3}
 \end{figure}

\begin{figure}
\caption{The fourth moment of the order-parameter as a 
          function of temperature for $\sigma=0.25$ and several 
system sizes.} 	
\label{fig-5}
 \end{figure}
 
\begin{figure}
          \caption{Logarithmic plot of the fourth moment of the 
order-parameter 
          as a function of system-size for $\sigma=0.25$ and several 
temperatures. The latter go between 0.1$J$ and 0.7$J$ in steps of 0.05$J$. The top curve corresponds to the lowest temperature.}        
 	\label{fig-7}
 \end{figure}
 
\begin{figure}
\caption{The correlation function exponent as a function of temperature for several values of $\sigma$.}        
\label{fig-eta}
\end{figure}

\begin{figure}
          \caption{The Binder function as a function of temperature 
          for different system sizes and several values of the 
          strength of the disorder in the strong disorder regime. (a)\  
$\sigma=0.49$, (b)\ $\sigma=0.64$, 
          (c)\ $\sigma=0.81$, and (d)\ $\sigma=1$. The solid lines are fits to the functional form described in the text.}
 	\label{fig-8}
 \end{figure}
 
 \begin{figure}
          \caption{The Binder function as a function of temperature 
          and disorder for several values of the 
          strength of the disorder and $L=16$. }
 	\label{fig-9}
 \end{figure}

\begin{figure}
          \caption{The universal scaling function in the disordered 
          phase as a function of $x=L/\xi$. }        
 	\label{fig-universal}
 \end{figure}

 \begin{figure}
\caption{Temperature dependence of the correlation length (in arbitrary units)  
in the strong disorder regime for the same values of $\sigma$ as in Fig.\ 9. The highest curve corresponds to the lowest value of the disorder. The full lines are derived from the fits in Fig.\ 8.}
\label{fig-10}
 \end{figure}

 \begin{figure}
\caption{The zero-temperature correlation length as a function of disorder. The unit of length is arbitrary. The solid line is a fit to the expression $\xi\sim \exp[b/(\sigma_c-\sigma)]$ with $b=0.69$.  The divergence at $\sigma_{c}$ signals the disorder-driven unbinding of vortices. }
\label{fig-11}
\end{figure}


\begin{references}
 	\bibitem{Rubinstein}  M. Rubinstein, B. Shraiman, and D. R. Nelson, 
 	Phys. Rev. B {\bf 27}, 1800 (1983).
 
 	\bibitem{ChaI}  M.-C. Cha, and H. A. Fertig, Phys. Rev. Lett. {\bf 
 	73}, 870 (1994).
 
 	\bibitem{GranatoI}  E. Granato, and J. M. Kosterlitz, Phys. Rev. 
 	B {\bf 33}, 6533 (1986).
 
 	\bibitem{GranatoII}  E. Granato, and J. M. Kosterlitz, Phys. Rev. 
Lett. {\bf 
 	62}, 823 (1989).
 
 	\bibitem{Fisher}  M. P. A. Fisher, T. A. Tokuyasu, and A. P. Young, 
Phys. Rev. Lett. {\bf 
 	66}, 2931 (1991).
 
 	\bibitem{Kosterlitz}  J. M. Kosterlitz, and D. J. Thouless, J. 
Phys. 
 	C {\bf 6}, 1181 (1973).
 
 	\bibitem{ForresterI}  M. G. Forrester, S. P. Benz, and C. J. Lobb, 
 	Phys. Rev. B {\bf 41}, 8749 (1989).
 
          \bibitem{Chakrabarti}  Amitabha Chakrabarti, and Chandan 
Dasguppta, 
 	Phys. Rev. B {\bf 37}, 7557 (1988).
 	
 	\bibitem{ForresterII}  M. G. Forrester, Hu Jong Lee, M. Tinkham, 
and C. J. Lobb, 
 	Phys. Rev. B {\bf 37}, 5966 (1988).
 
 
 	\bibitem{ChaII}   M.-C. Cha, and H. A. Fertig, Phys. Rev. Lett. 
{\bf 
 	74}, 4867 (1995).
 
 	\bibitem{Natterman} T. Natterman, S. Scheidl, S. E. Korshunov, and 
M. 
 	S. Li, J. Phys. I (France) {\bf 5}, 565 (1995).  
 	
 	\bibitem{Ozeki}  Y. Ozeki, and H. Nishimori, J. Phys. A {\bf 26}, 
 	3399 (1993).

	\bibitem{Maucourt} J. Maucourt, and D. R. Grempel, umpublished.
 	
	\bibitem{bhatt} R. N. Bhatt, and A. P. Young, Phys. Rev. B {\bf 37}, 
5606 (1988).

 	\bibitem{ray} P. Ray, and M. A. Moore, Phys. Rev. B {\bf 45}, 5361 
 	(1992). 
 
 	\bibitem{Binder}  K. Binder,Z. Phys. B {\bf 43}, 119 (1981)
 
 \end{references}
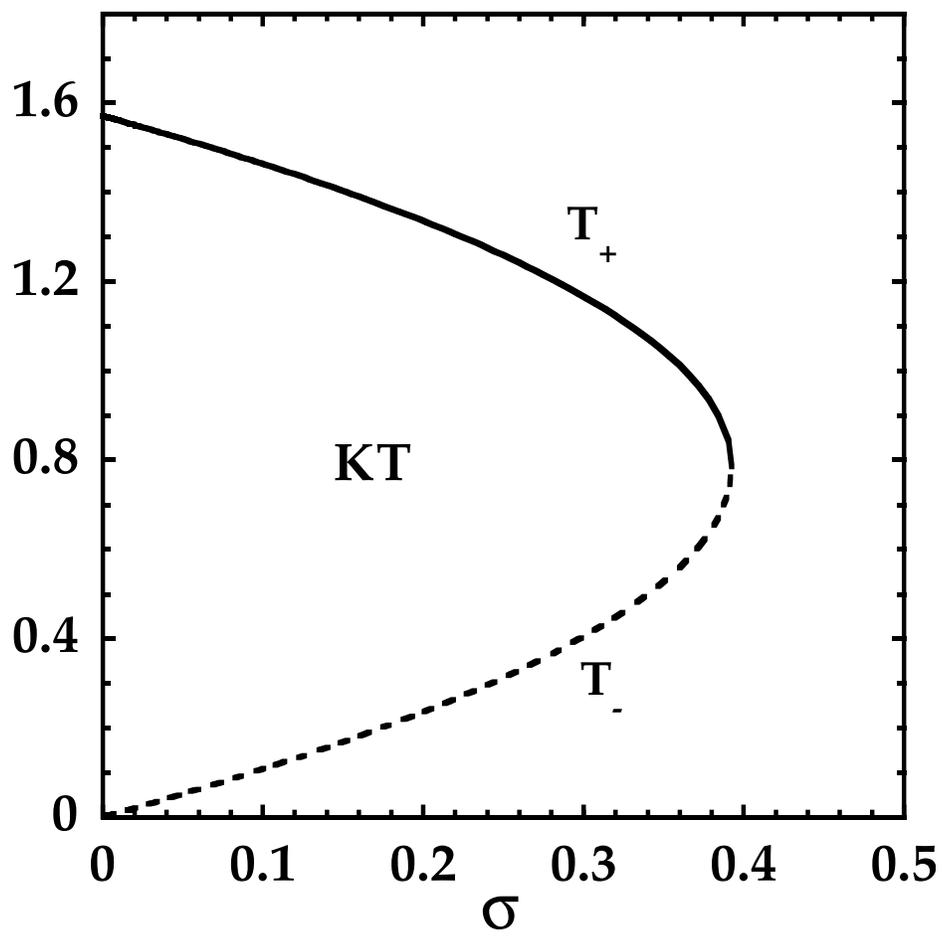
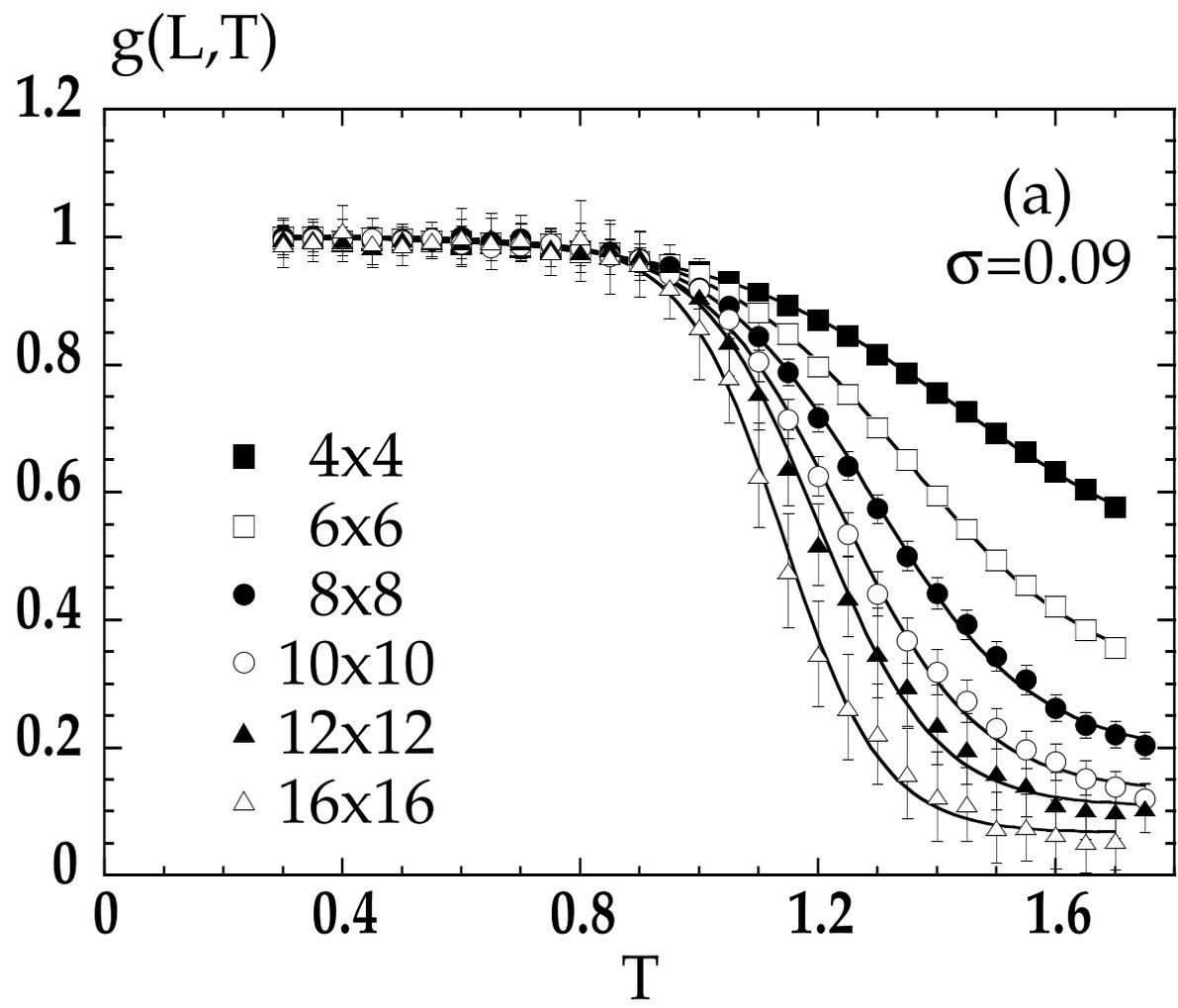
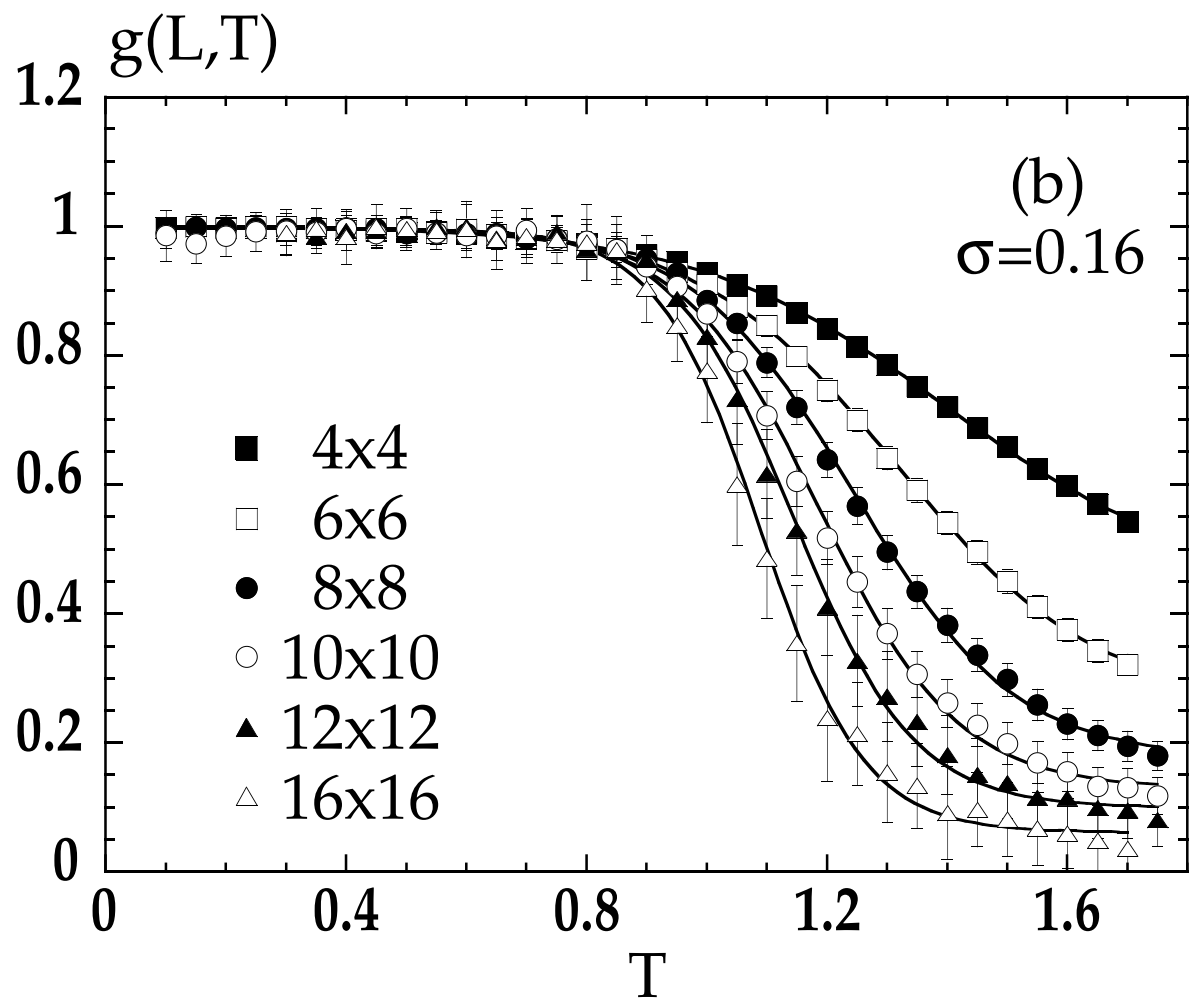
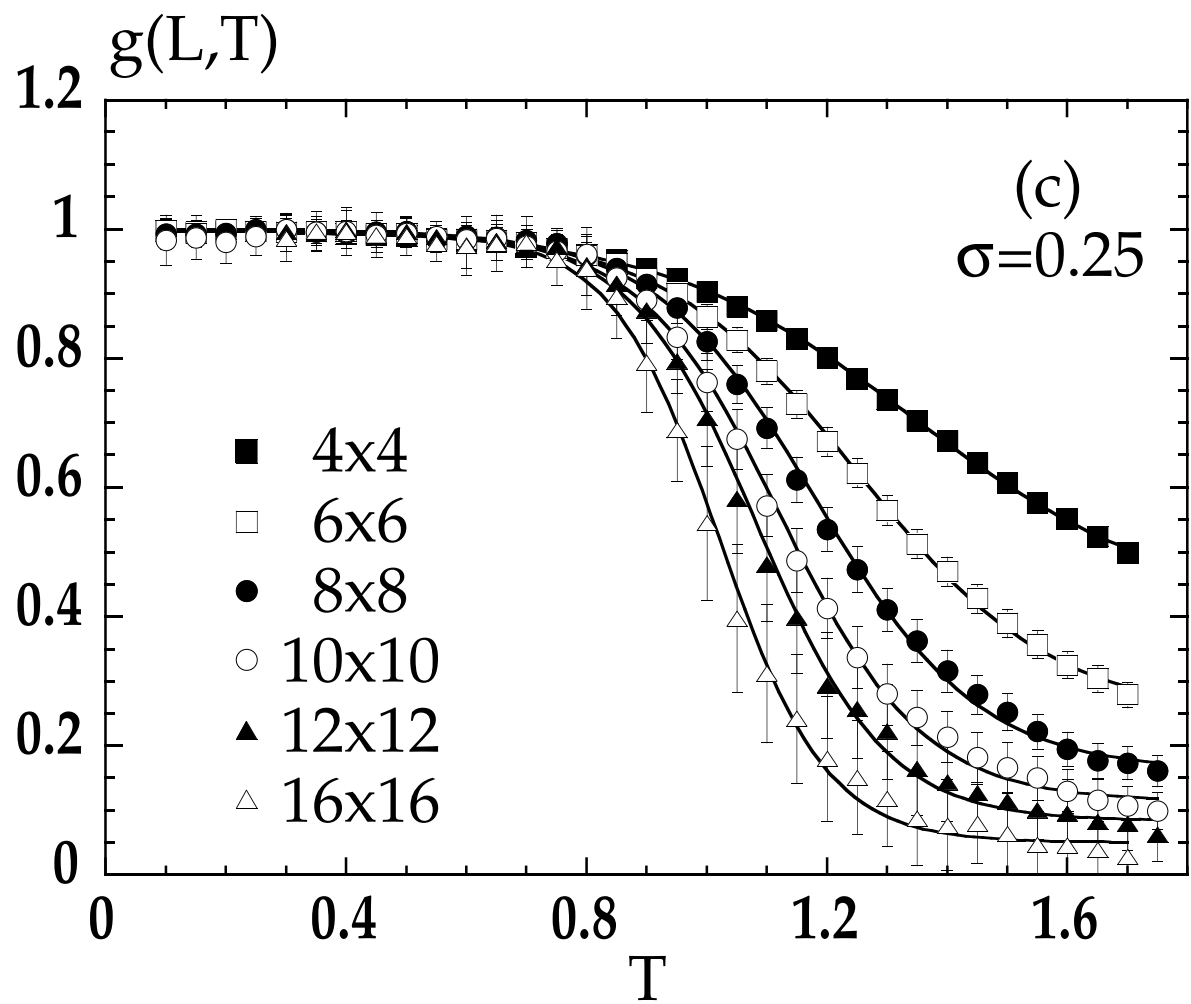
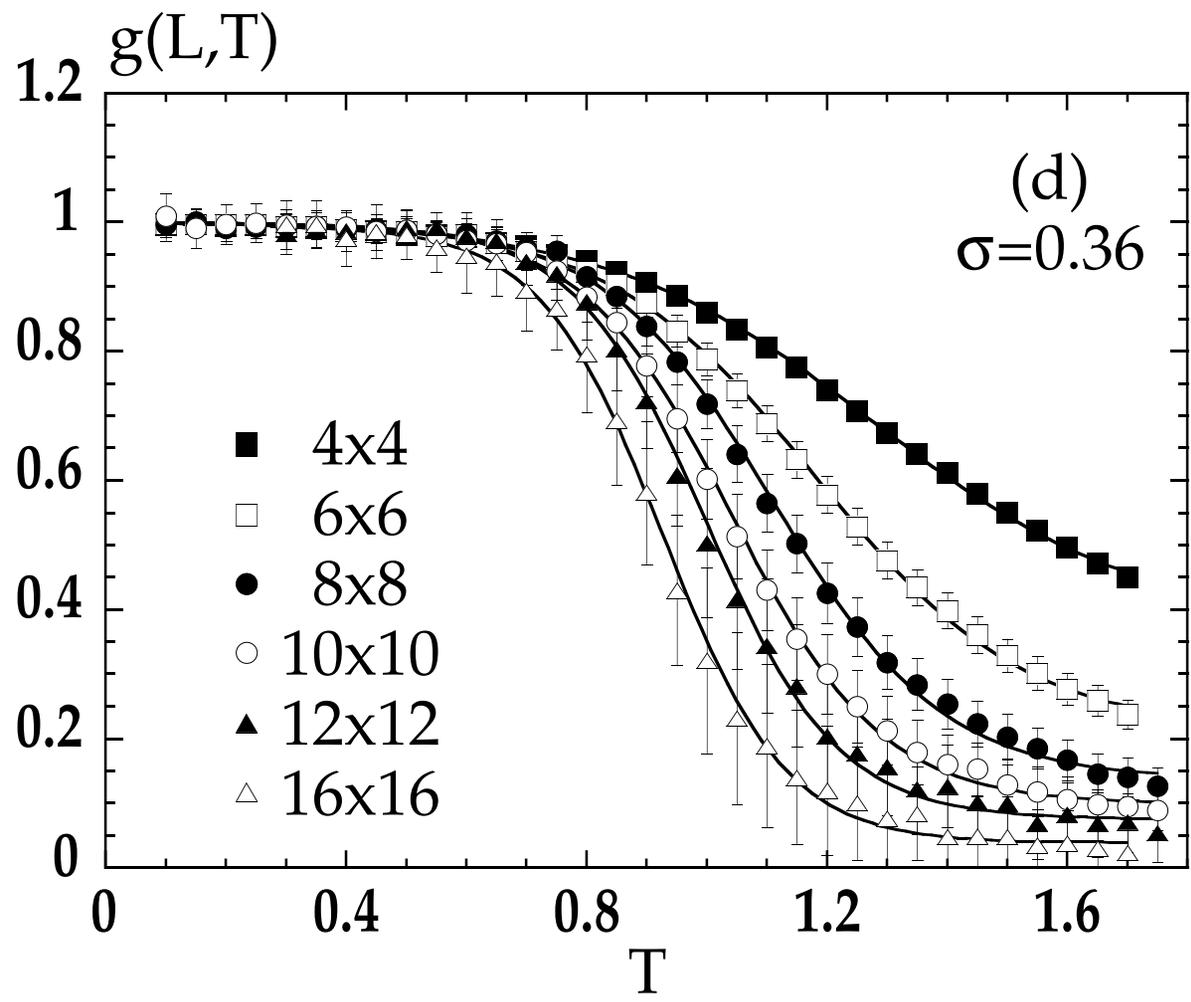
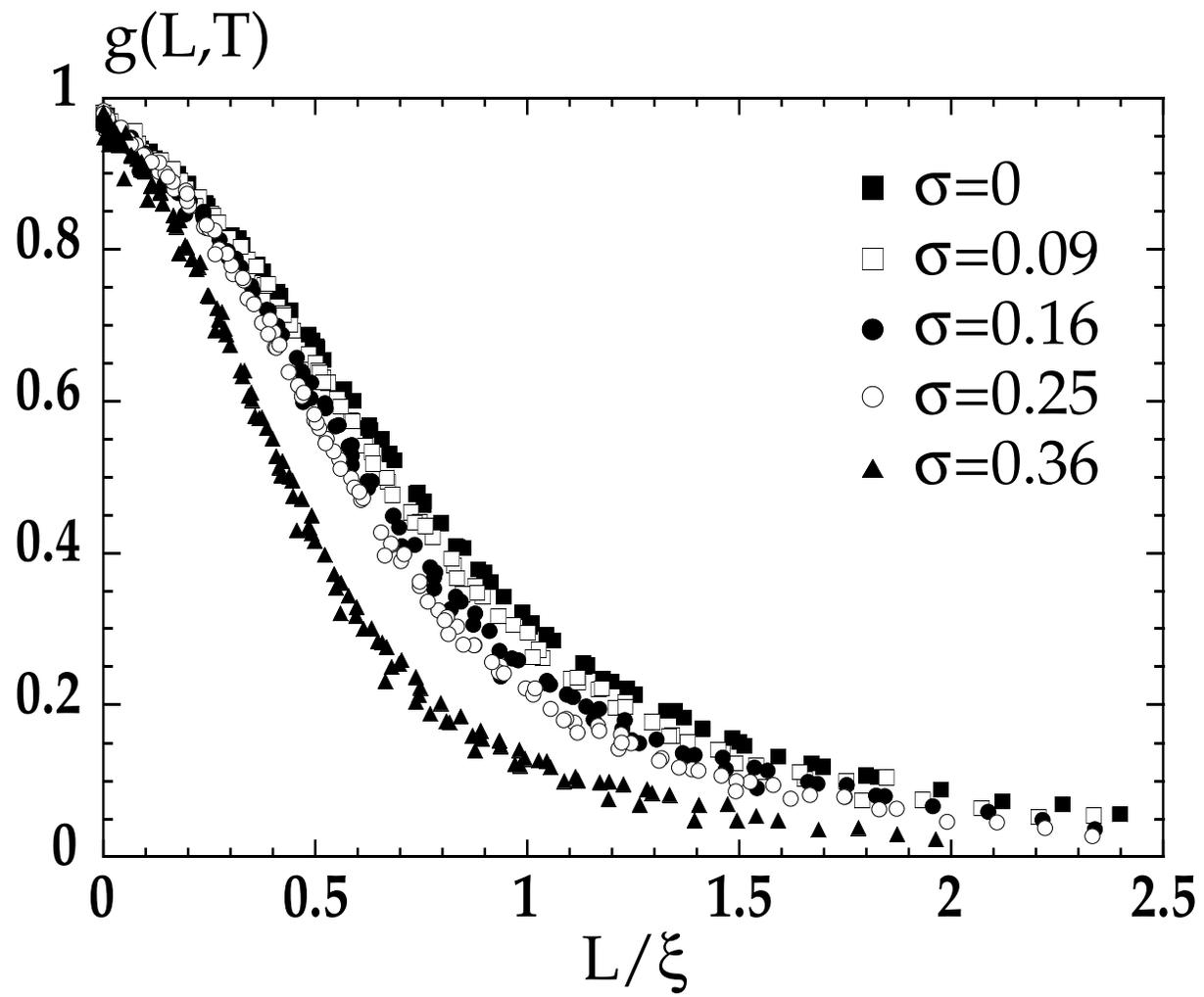
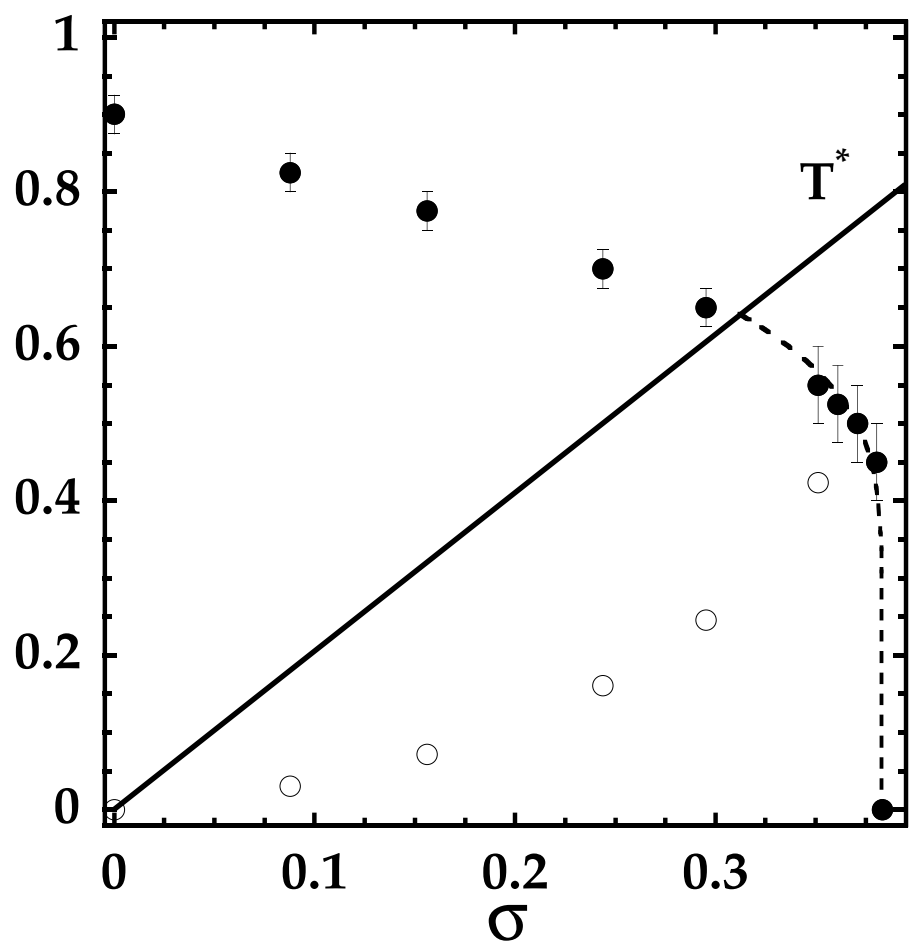
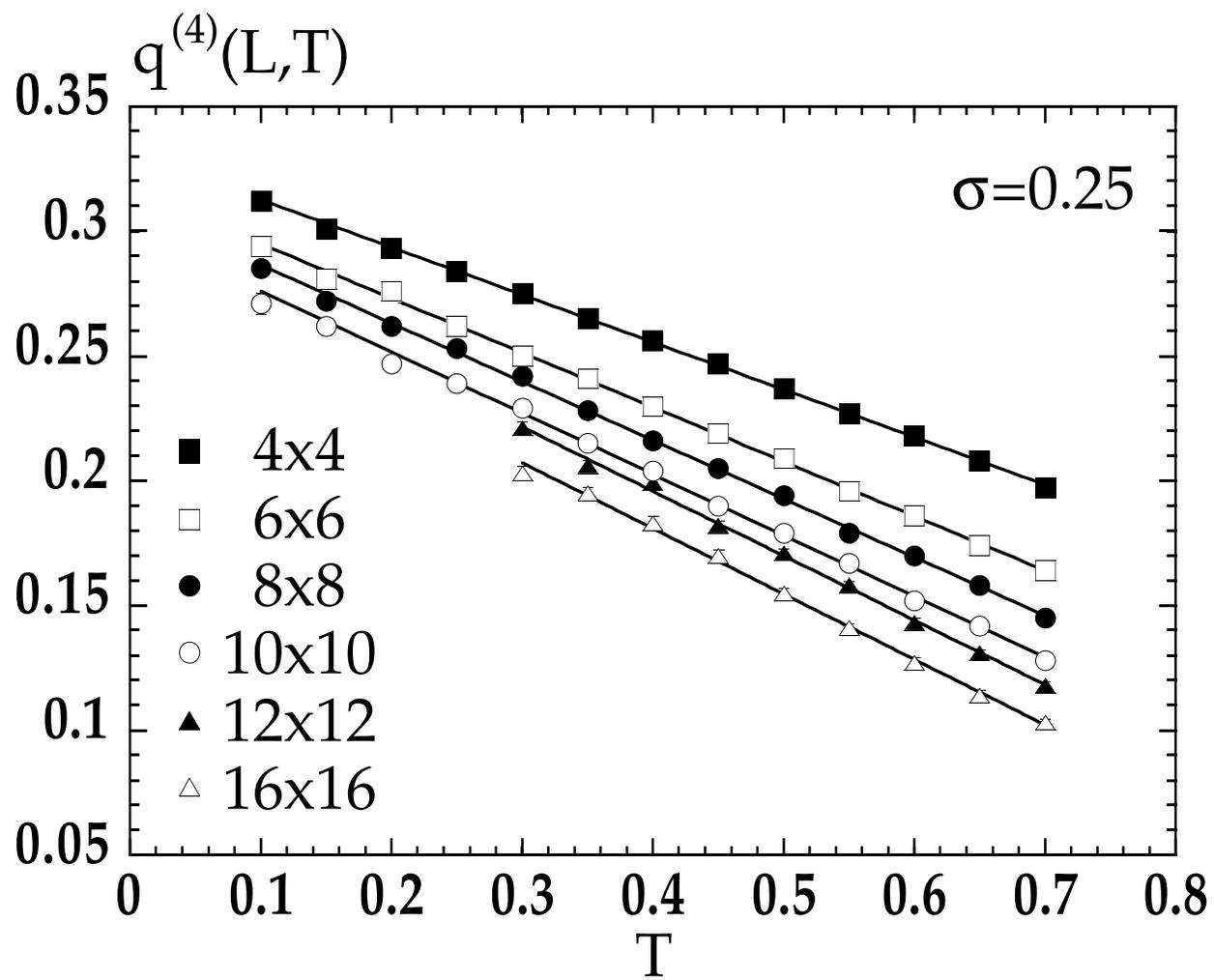
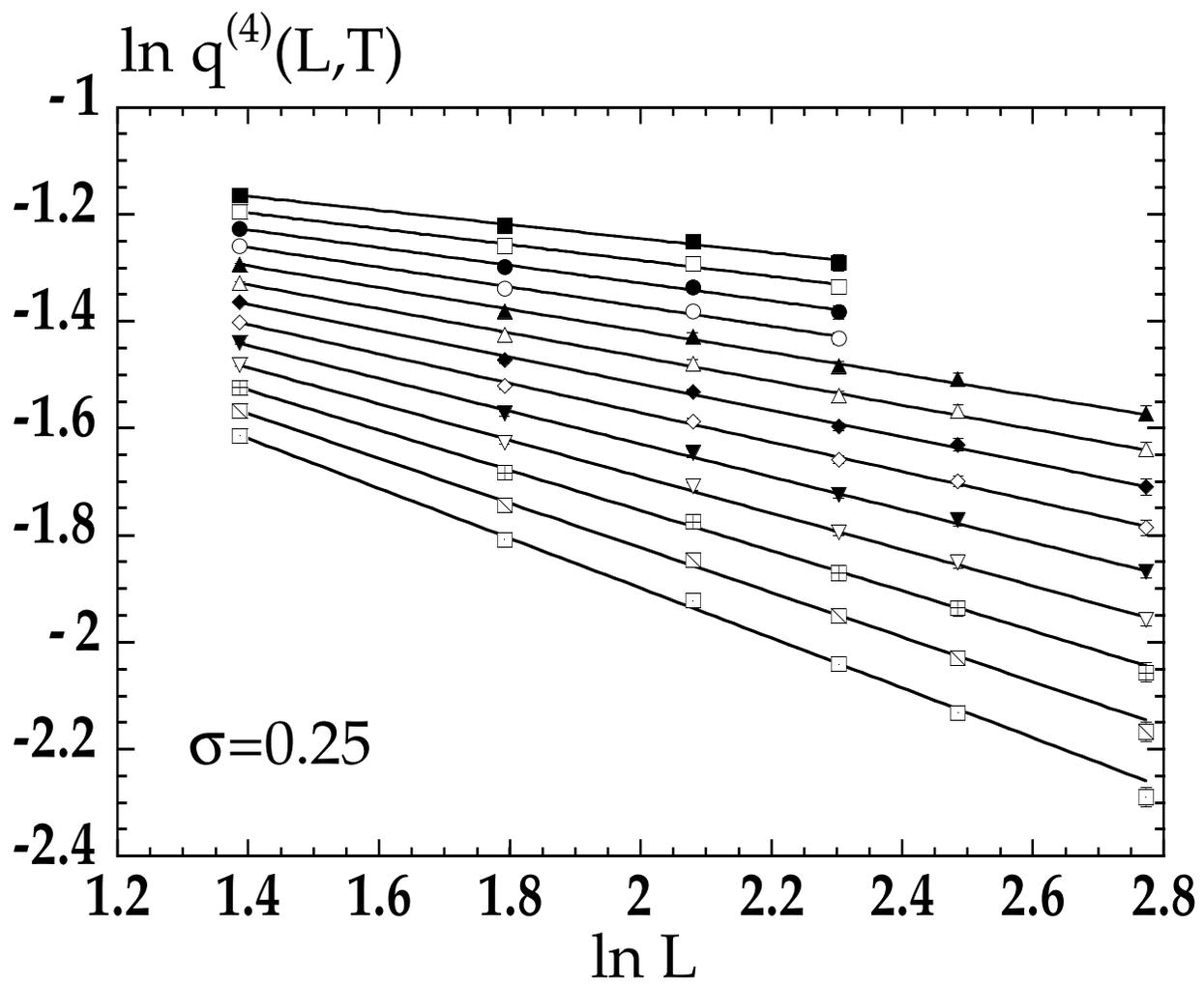
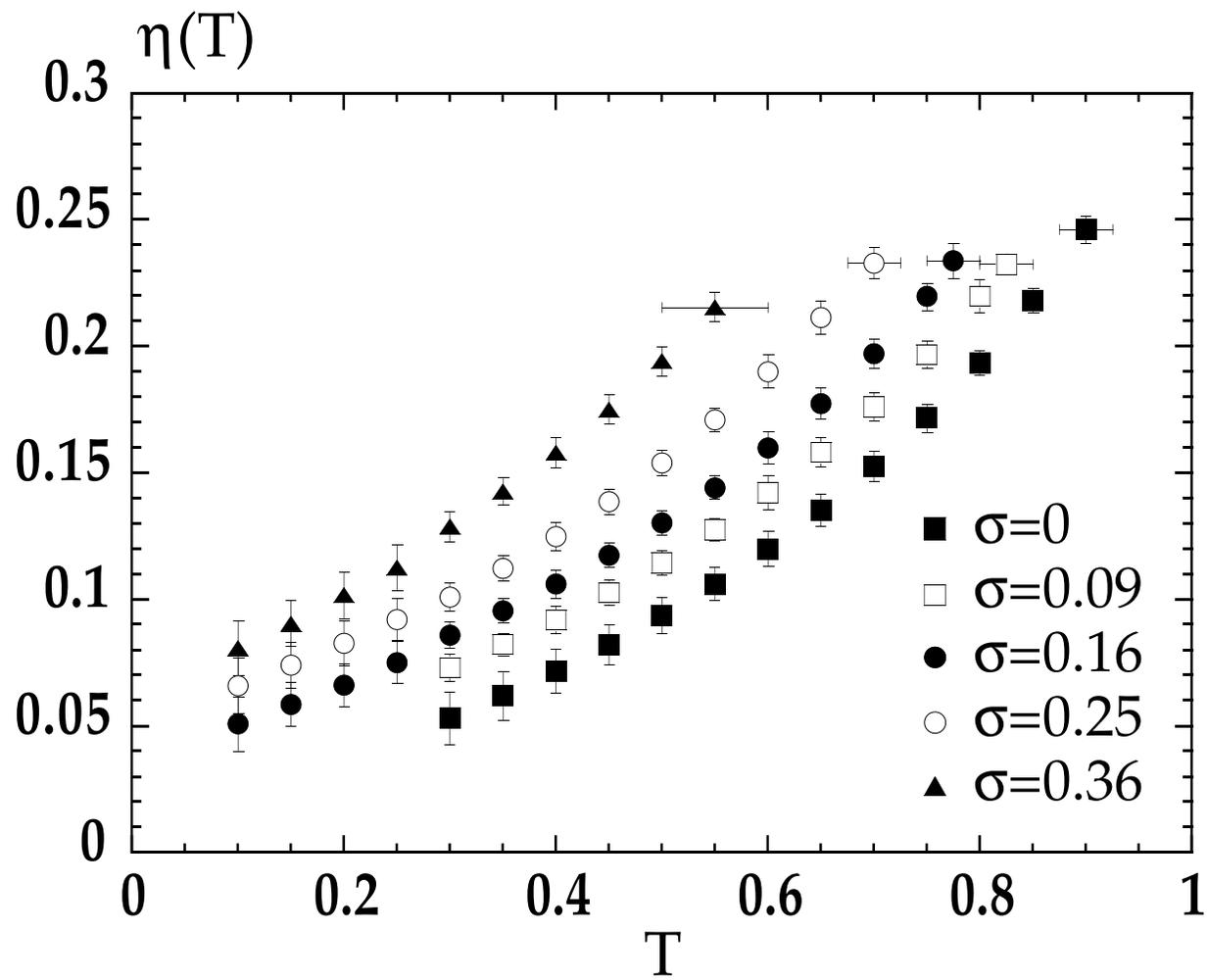
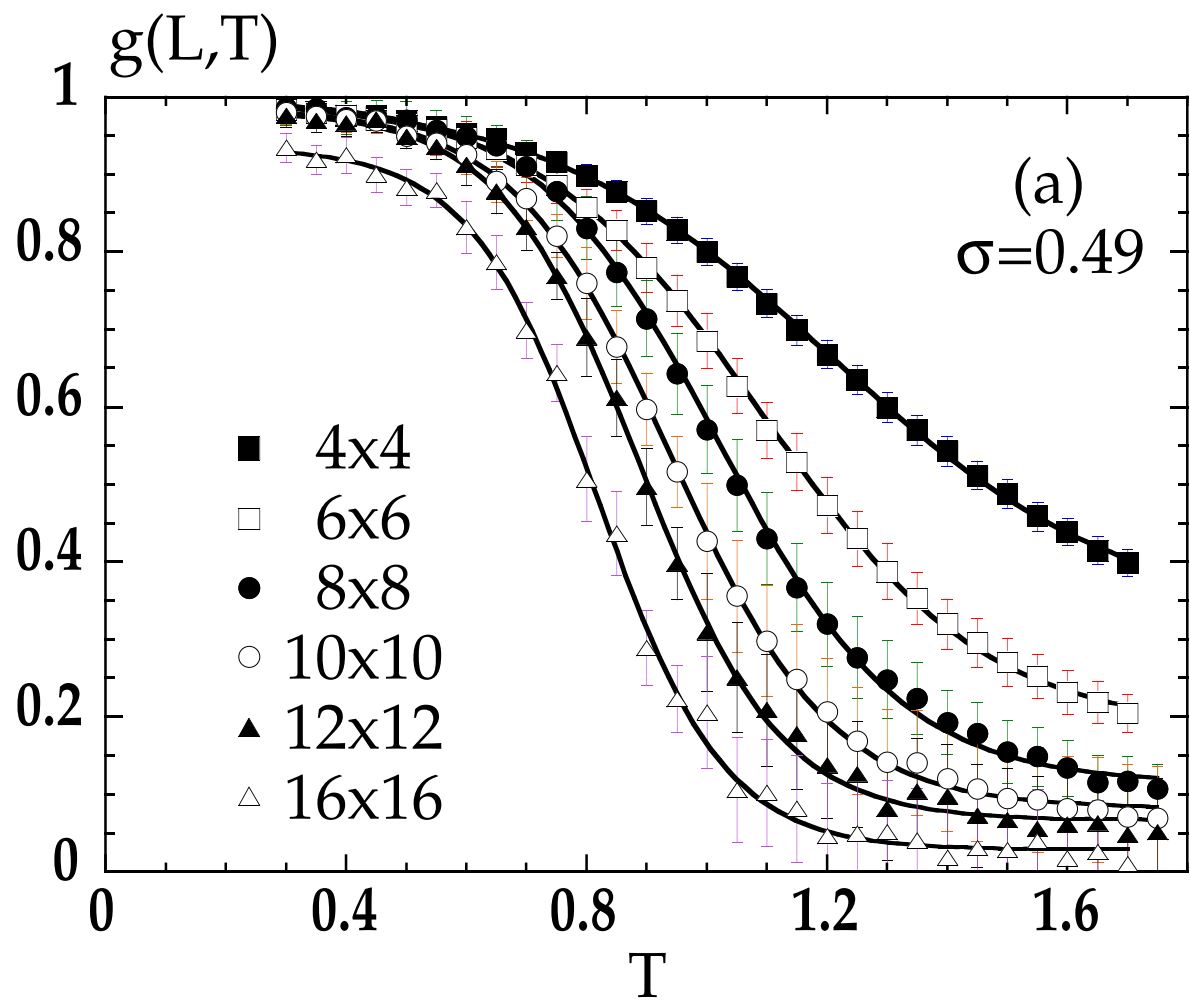
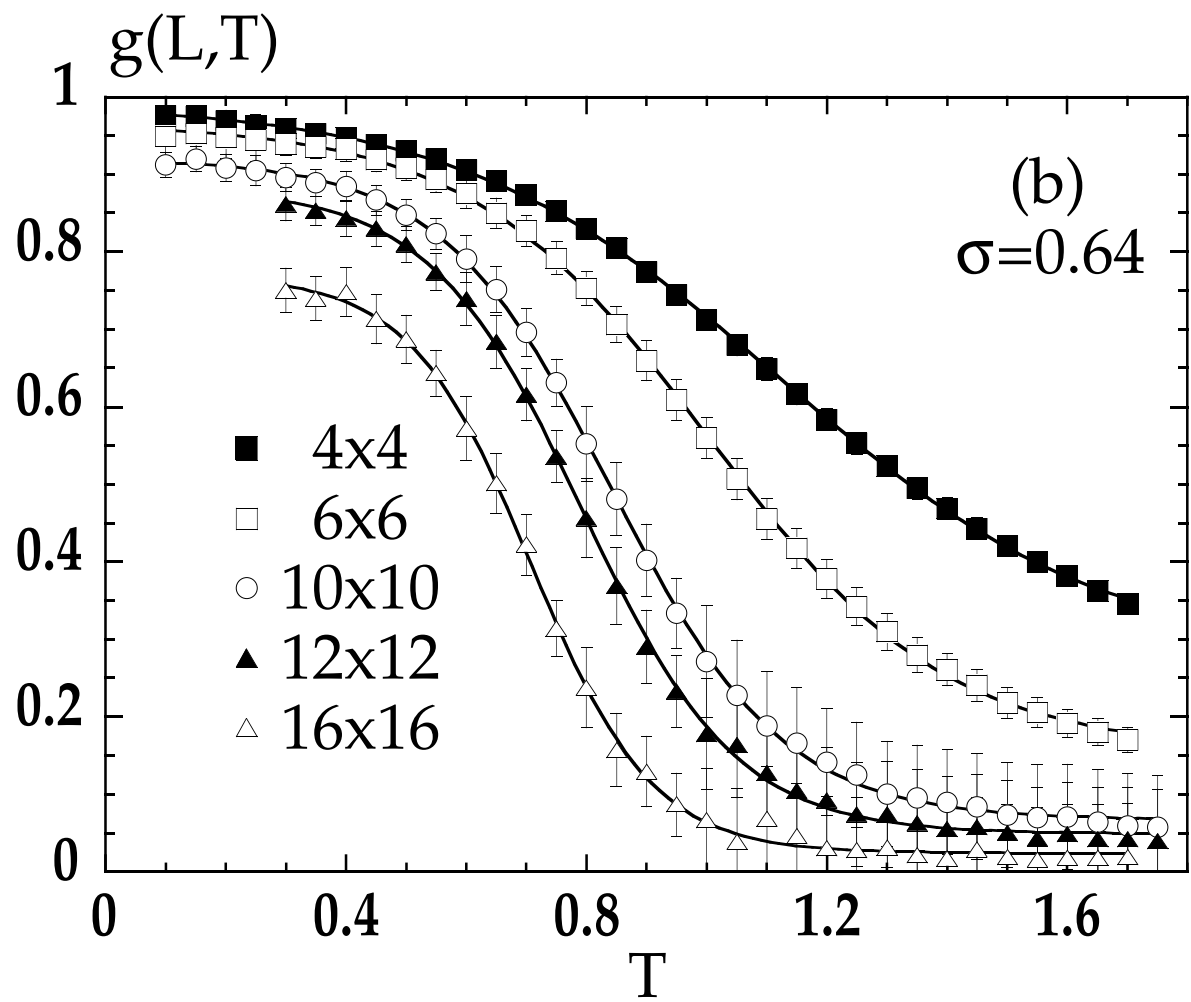
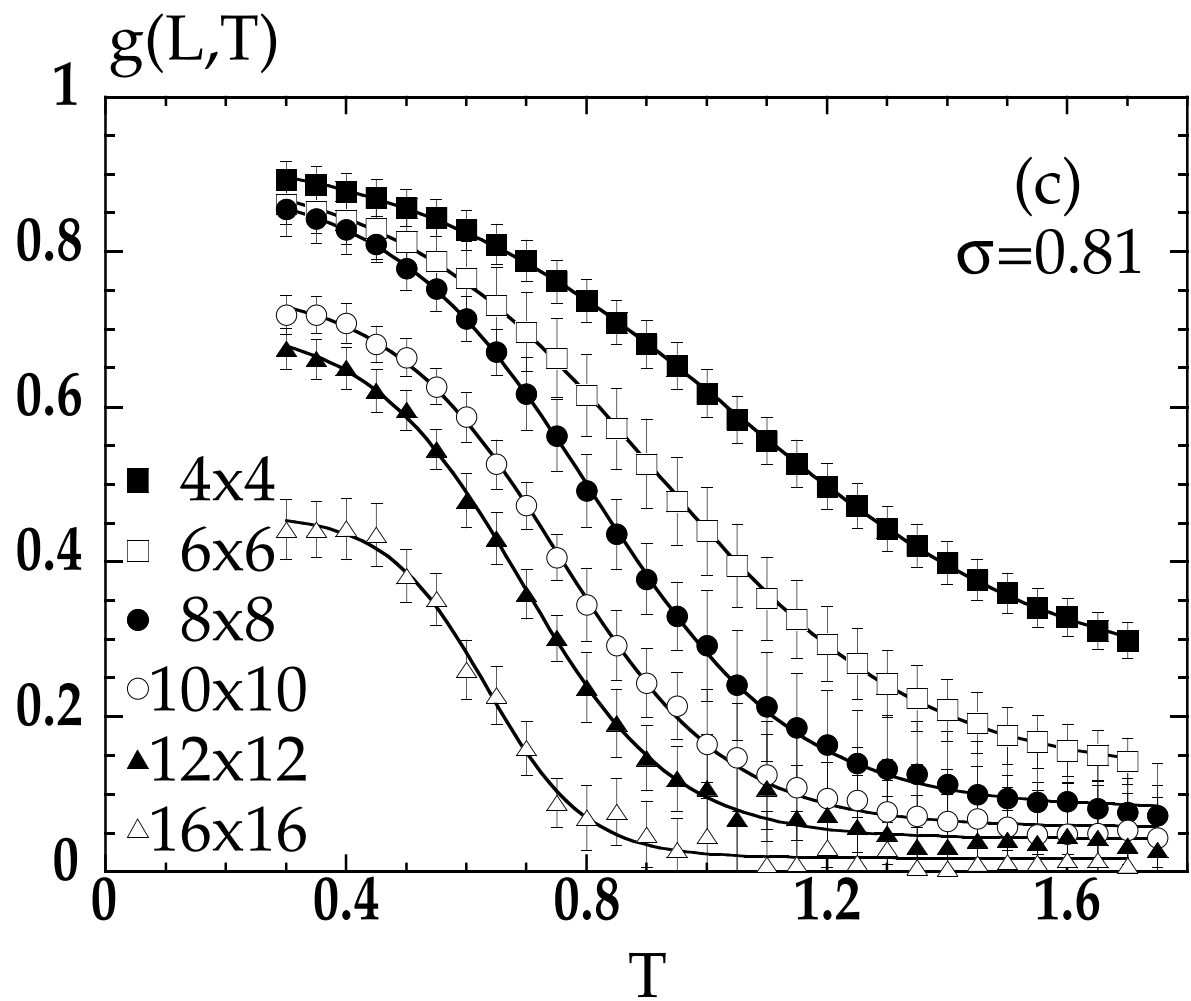
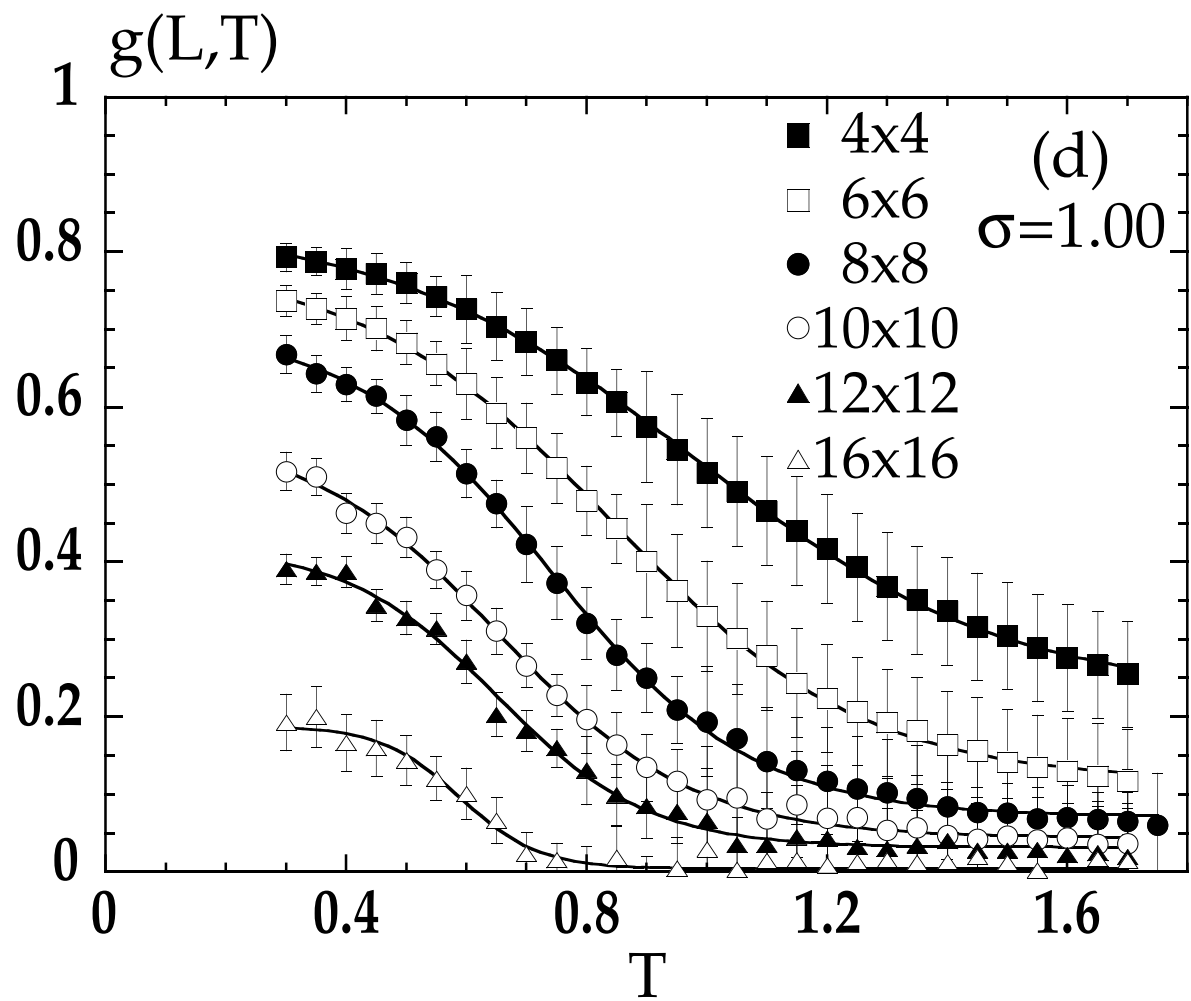
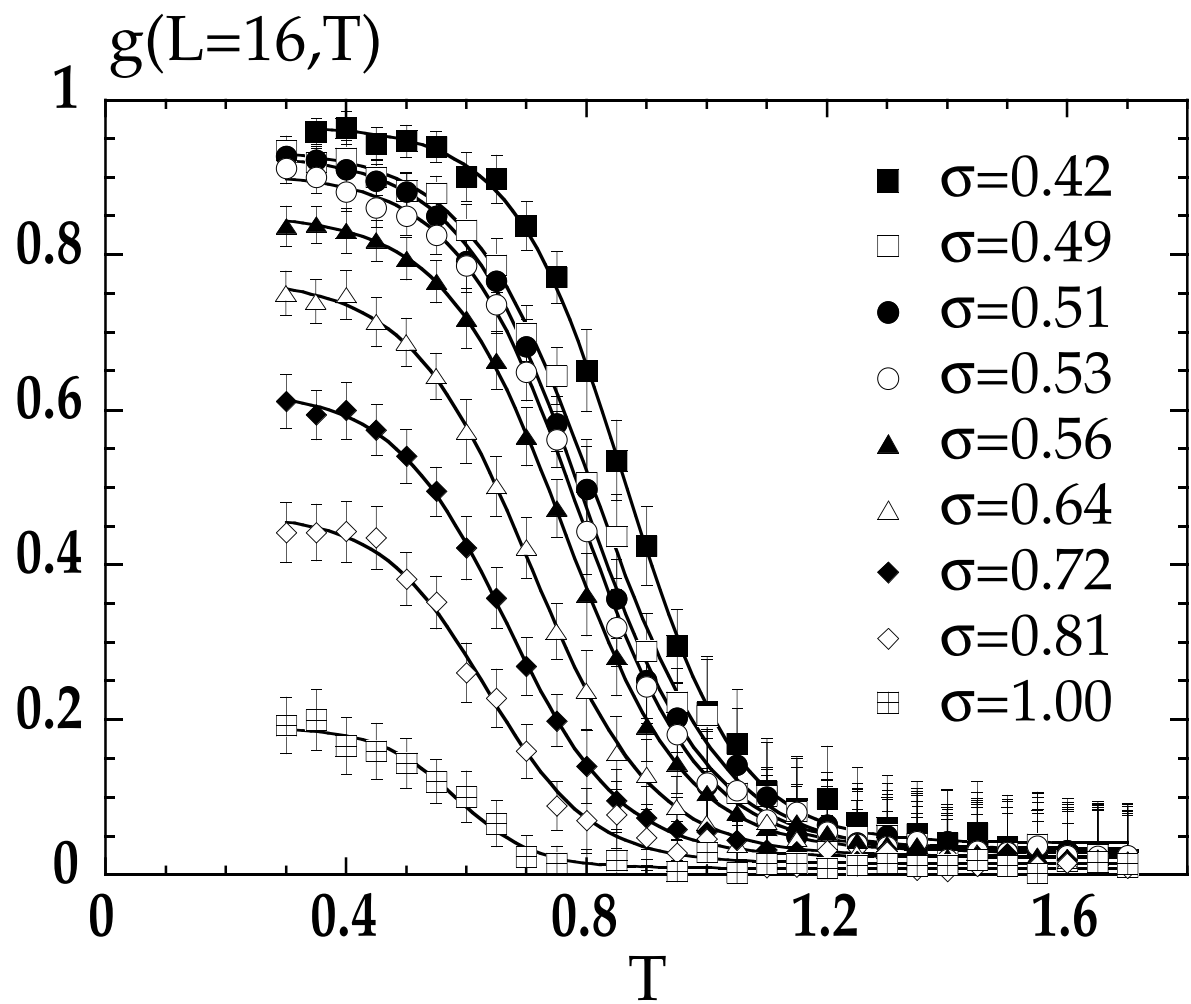
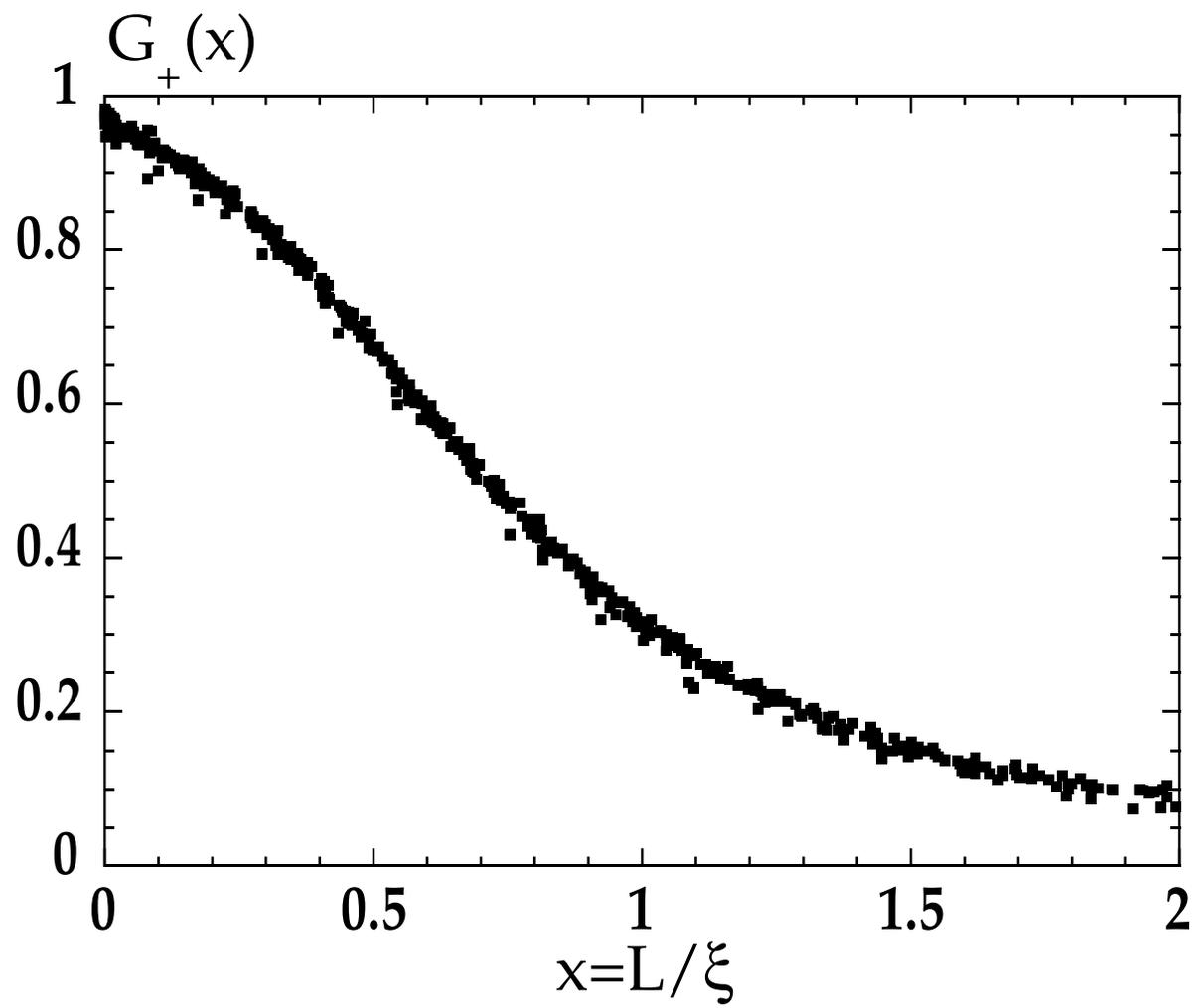
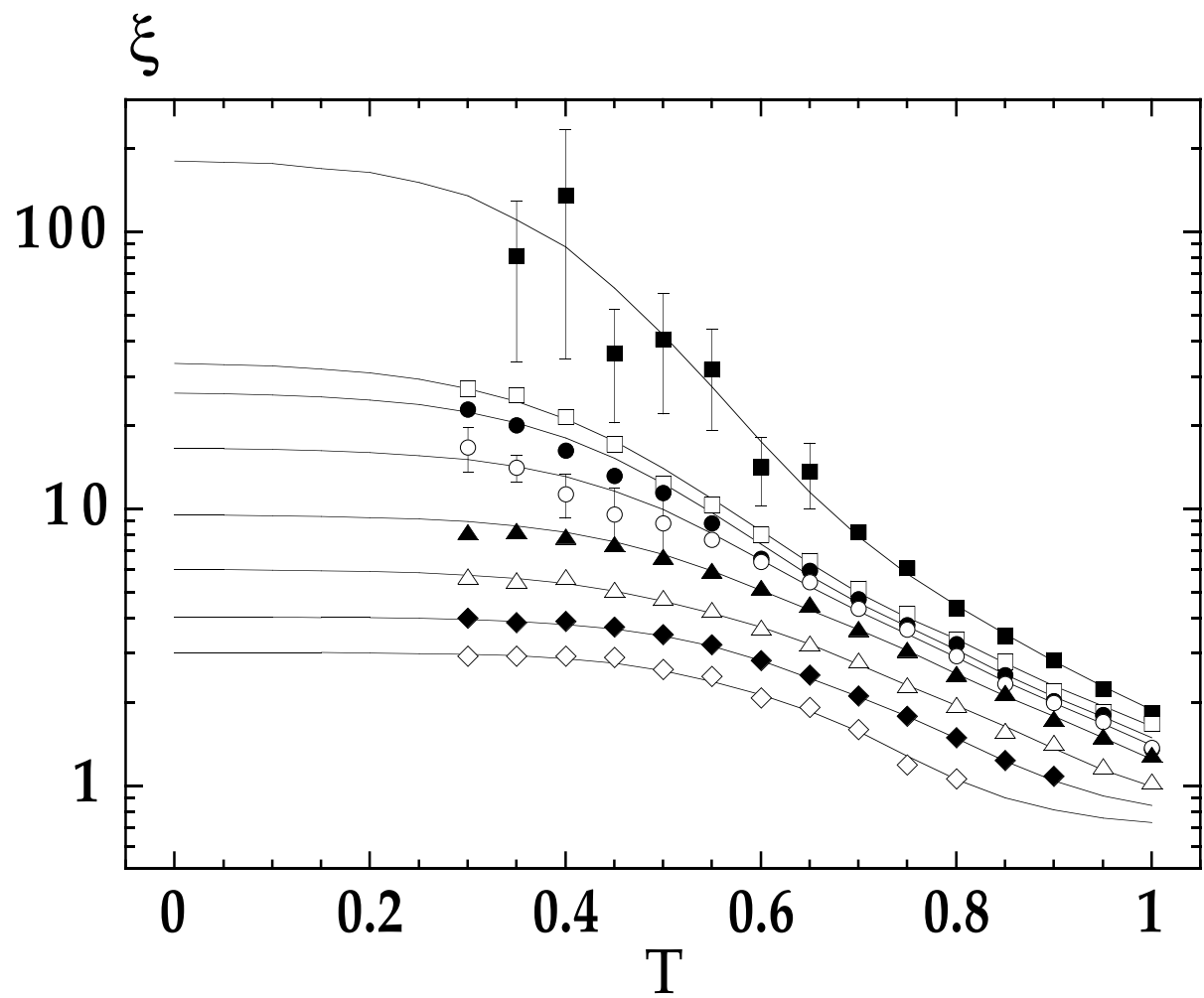
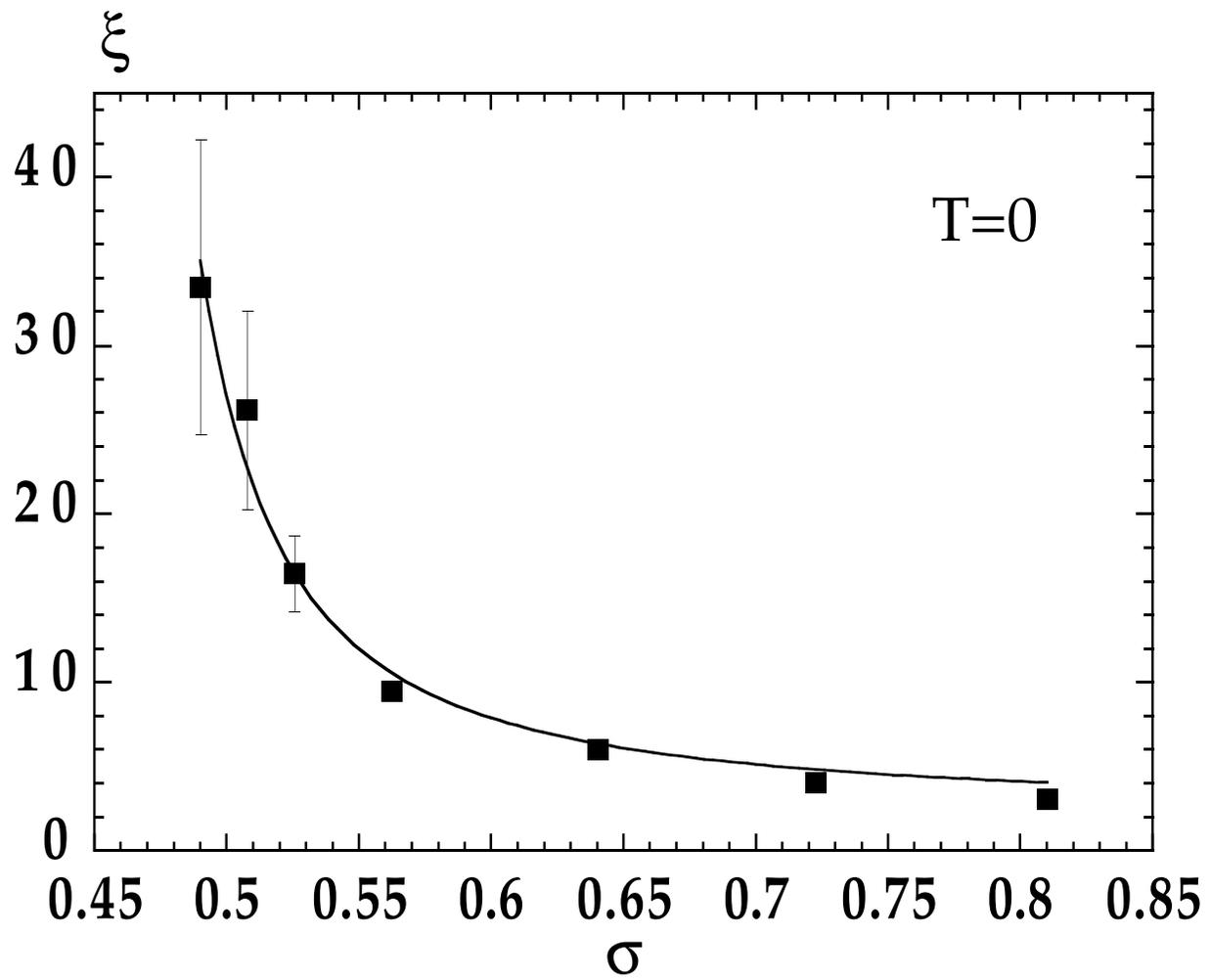
\end{document}